\documentclass[twocolumn,showpacs,preprintnumbers,amsmath,amssymb]{revtex4}
\usepackage{amsmath,amssymb}
\usepackage{graphicx}% Include figure files
\usepackage{dcolumn}% Align table columns on decimal point
\usepackage{bm}% bold math
\usepackage{mathptmx}

\def\Beq{\begin{equation}}
\def\Eeq{\end{equation}}

\def\Bea{\begin{eqnarray}}
\def\Eea{\end{eqnarray}}

\def\Beaa{\begin{eqnarray*}}
\def\Eeaa{\end{eqnarray*}}

\def\BD{\begin{description}}
\def\ED{\end{description}}

\def\BC{\begin{center}}
\def\EC{\end{center}}

\def\del{\partial}
\def\l<{\langle}
\def\r>{\rangle}
\def\({\left(}
\def\){\right)}

\def\lra{\longrightarrow}

\def\bmD{\bm{D}}
\def\mrE{\mathrm{e}}
\def\nb{\nabla}
\def\nbbr{\breve{\nb}}
\def\Dot{ {\> {}_ \centerdot}}  
\def\gddot{\g_{{}^{\, {}_ \centerdot} {}^{\, {}_ \centerdot}}}
\def\half{\frac{1}{2}}
\def\ddt{\frac{d}{d\t}}
\def\dddtt{\frac{d^2}{d\t^2}}
\def\a{\alpha}
\def\b{\beta}
\def\g{\gamma}
\def\G{\Gamma}
\def\d{\delta}
\def\D{\Delta}

\def\l{\lambda}

\def\th{\theta}
\def\Th{\Theta}
\def\om{\omega}
\def\Om{\Omega}

\def\s{\sigma}
\def\t{\tau}
\def\^{\wedge}

\def\C{\mathcal{C}}

\def\M{\mathcal{M}}

\def\v0{\vec{0}}

\def\ol{\overline}
\def\ul{\underline}

\begin{document}

%\setlength{\baselineskip}{1cm}

%\maketitle
%%%%%%%%%%%%%%%%%%%%%%%%%%%%%%%%%%%%%%%%%%%%%%%%%%%%%%%%%%%%%%%%%
%%%%%%%%%%%%%%%%%%%%%%%%%%%%%%%%%%%%%%%%%%%%%%%%%%%%%%%%%%%%%%%%%
%Body of Paper
%%%%%%%%%%%%%%%%%%%%%%%%%%%%%%%%%%%%%%%%%%%%%%%%%%%%%%%%%%%%%%%%%
%%%%%%%%%%%%%%%%%%%%%%%%%%%%%%%%%%%%%%%%%%%%%%%%%%%%%%%%%%%%%%%%%

\title{Influence of geometry variations on the gravitational focusing of timelike geodesic
 congruences}% Force line breaks with \\

\author{Masafumi Seriu}
 \email{mseriu@edu00.f-edu.u-fukui.ac.jp}
 \affiliation{%
Mathematical and Quantum Science Group, 
Division of  Physical Science, Graduate School of Technology and Engineering, 
University of Fukui, 
Fukui 910-8507, Japan\\}%

%\date{\today}% It is always \today, today,
             %  but any date may be explicitly specified

%%%%%%%%%%%%%%%%%%%%%%%%%%%%%%%%%%%%%%%%%%%%%%%%%%%%%%%%%%%%%%%%%
%%%%%%%%%%%%%%%%%%%%%%%%%%%%%%%%%%%%%%%%%%%%%%%%%%%%%%%%%%%%%%%%%
%ABSTRACT
%%%%%%%%%%%%%%%%%%%%%%%%%%%%%%%%%%%%%%%%%%%%%%%%%%%%%%%%%%%%%%%%%
%%%%%%%%%%%%%%%%%%%%%%%%%%%%%%%%%%%%%%%%%%%%%%%%%%%%%%%%%%%%%%%%%
\begin{abstract}
We derive a set of equations describing the linear response of 
 the convergence properties of a geodesic congruence 
to  arbitrary geometry variations. 
It is a combination of 
equations describing 
the deviations from the standard Raychaudhuri-type equations 
due to the geodesic shifts and 
an equation describing the geodesic shifts due to 
the geometry variations. 
In this framework,  the geometry variations, which can be chosen arbitrarily, 
serve as  probes to investigate the gravitational contraction processes from 
various angles. 

We apply the obtained framework to the case of conformal geometry variations, 
characterized by an arbitrary  function $f(x)$, and see that 
  the formulas get simplified to a great extent. 
We investigate the response of 
the convergence properties of geodesics 
in the latest phase of gravitational contractions by 
restricting the class of conformal geometry variations 
to the one satisfying the strong energy condition. 
We then find  out that 
in the final stage,  $f$ and $\bmD  \cdot \bmD f$ control the overall contraction 
behavior  and 
that the contraction rate gets larger
when $f$ is negative and $|f|$ is  so large as  to overwhelm  
$|\bmD  \cdot \bmD f|$. (Here $\bmD  \cdot \bmD$ is the  Laplacian operator on 
the spatial hypersurfaces orthogonal to the geodesic congruence in concern.) 

To get more concrete insights, we also apply the framework to the time-reversed 
Friedmann-Robertson-Walker model as the simplest case of the singularity formations.
  \end{abstract}


%\keywords{Suggested keywords}%Use showkeys class option if keyword
                              %display desired
%\thanks{}

\maketitle

\section{\label{sec:Introduction}Introduction}

It has been known that the spacetime manifolds with high symmetries often 
contain singularities.  Typical examples are 
 the initial big-bang  singularity in the spatially homogeneous  and isotropic 
universe models  and the curvature singularities in the  black-hole 
solutions.  
The singularity theorems~\cite{HawkingPenrose,HawkingEllis,Wald,Joshi} have shown that, however,  
the spacetime singularities do not result from high symmetries, but 
are quite  general features of spacetimes satisfying  
reasonable physical conditions,
 such as the energy condition and no closed timelike curves.  

Though the singularity theorems show  the generality of  singularities 
in physically reasonable spacetimes,  it is also true that 
their statements are too universal and general to get detailed information on  
 gravitational collapses.  

Among these processes needed to be clarified, 
the black-hole formations are especially important ones.  
Though the spacetime structures {\it after} black-hole formations 
are quite well-understood by the uniqueness theorem for the Kerr 
solution and its related theorems~\cite{HawkingEllis, Wald}, 
little is clarified about the black-hole formation processes themselves. 
Indeed,  the cosmic censorship hypothesis~\cite{Penrose1,Penrose2} and the 
hoop conjecture~\cite{hoop}, which are the  
two central conjectures for the black-hole formations, have not been satisfactorily  
proved so far even though there have been no physically reasonable model found 
 manifestly contradicting with these conjectures~\cite{Seriu}. 
One of the reasons for this situation might be the fact that 
there is no established framework for analytically describing the black-hole formation 
processes. 

Here let us pay attention to the Raychaudhuri 
equation~\cite{HawkingEllis,Wald,Raych1,Raych2,KarSengupta}, which describes 
the focusing property of a given geodesic congruence. The equation 
indicates that,
 once the expansion $\th$ along a geodesic in the congruence gets negative, 
it approaches to $-\infty$ within some finite proper-time along the geodesic provided 
that the strong energy condition is satisfied (see Sec.\ref{sec:Focusing}). 
This phenomenon signals  
the occurrence of the conjugate point (the focal point) in the  future which in turn 
implies the singularity (in the sense of the timelike geodesic incompleteness) 
in the future within a finite proper-time provided some other 
conditions are also satisfied~\cite{HawkingEllis,Wald,Joshi}.  

Considering the above fact, we here choose the strategy to  pay attention to the 
convergence properties of the geodesic congruence as the 
starting point for the analytical description of the gravitational contractions. 
More specifically, we here aim at constructing a theoretical framework  
describing how the convergence properties of a given 
 time-like geodesic congruence are influenced   by 
the slight variation in  geometrical properties around the geodesic congruence. 
In other words, we study the linear response of the 
convergence properties of the geodesic congruence to the arbitrary 
geometry variations.
Here the origins of the geometry variations are not specified and 
they can either be  real  physical processes  or    virtual displacements.   
Since the geometry variations can be arbitrarily chosen by hand, then, we might be 
able to  use them as probes to investigate the gravitational contraction processes from various angles.

We now show  the  plan for constructing  the framework. 
The outline of the physical process in question is as follows: 
When  the geometry is varied, geodesics are also shifted accordingly, which in turn 
changes the convergence properties of the geodesics. 
The construction of the framework, thus, naturally  consists of two steps. 
As the first step, 
 we establish a key equation relating the geometry variation to the geodesic shift. 
As the second step, then,   
we investigate the variations in the convergence properties of 
the geodesic congruence caused by these geodesic shifts.
Combining these two steps, we can construct a framework which describes
 the changes in the convergence properties of the geodesic congruence 
due to the geometry variations.

One key point of the above plan   resides in  the first step. 
Since the geodesic is the integral curve of a tangent vector field, it is 
a global object, which clearly causes some difficulty in 
pursuing our plan.  
We shall tackle this problem by introducing a suitable  vector describing 
the geodesic shift.  
Furthermore it turns out that 
using solely the  component of this geodesic shift vector orthogonal to the geodesic is 
essential, making the expressions much simpler as well as 
giving much more transparent interpretations of the equations. 
In this way, we shall  find out the  key equation. 

Another key point is to  introduce  a 1-parameter family of geometries on a fixed manifold $\M$, by means of which 
we can mathematically handle the geometry variations and derive a set of 
equations describing their influence on 
the convergence properties of a given geodesic congruence.  

After establishing general formulas, 
we shall then 
pay special attention to  the case of the {\it conformal} geometry variations.  
This case not only makes all the expressions much simpler, but also 
is quite important for several  applications.  
We shall  analyze  the influence on the convergence properties 
of geodesics in the final phase of gravitational contractions by 
restricting the class of conformal geometry variations 
to the one satisfying the strong energy condition. 
We then find  out that,  
when $f(x)$ is the function characterizing the conformal geometry variation, 
 $f$ and $\bmD  \cdot \bmD f$ control the overall contraction 
behavior in the final stage (where $\bmD  \cdot \bmD$ is the  Laplacian operator 
induced on the spatial hypersurfaces orthogonal to the timelike geodesic congruence 
in question),  and that the contraction rate gets larger
when $f$ is negative and  $|f|$ is  so large as  to overwhelm  
$|\bmD  \cdot \bmD f|$.

This paper is organized as follows. In Sec.\ref{sec:Basics}, we shall review  
the established results regarding the deformation properties of  timelike geodesic 
congruences needed in the paper. 
The basic quantities of our analysis - the expansion $\th$, the shear $\s_{ab}$  and the twist $\om_{ab}$-  
and the Raychaudhuri-type equations  shall be  introduced  there. 
This section is also served to fix the notations and the definitions of the terms 
 adopted throughout the  paper. 
The main parts 
of the paper are 
Sec.\ref{sec:Deviation} and Sec.\ref{sec:Convergence_shift}.   
In  Sec.\ref{sec:Deviation}, we shall derive a key equation 
describing the geodesic shift  caused by  a geometry variation. 
In Sec.\ref{sec:Convergence_shift}, then, we shall derive a system of equations 
describing changes in the deformation properties of the geodesic congruences 
due to geometry variations. 
In Sec.\ref{sec:Conformal},  we shall focus on the case of the conformal variations 
and shall see all the formulas reduce to much simpler counterparts in this case. 
We shall then apply our framework to one concrete, explicitly analyzable model;
the conformal variations applied to  the time-reversed Friedmann-Robertson-Walker 
model.  
Based on the results of the previous section,   we shall analyze 
in Sec.\ref{sec:Focusing} the changes in the contraction properties 
due to the conformal geometry variations in some detail. 
Section \ref{sec:Summary} is devoted for the summary and several discussions.

\section{\label{sec:Basics}Basic formulas}

\subsection{\label{subsec:congruence}Timelike geodesic congruences}
Let $(\M, g)$ be a smooth $n$-dimensional  spacetime manifold with 
 signature $(- + + \cdots +)$.
 
Let $\nb_a$ be the standard  covariant derivative on $(\M, g)$, 
satisfying the metricity condition  ($\nb_a g_{bc}=0$)  and   
the torsion-free condition ($(\nb_a \nb_b - \nb_b \nb_a)  f =0$ for any 
smooth scalar function $f$).

From now on, any  timelike geodesic $\g$ is assumed to be affine-parametrized 
by the proper-time $\t$ along it. Thus the  tangent vector of $\g$, 
${\xi}^a (\t) \equiv \(\del/\del \t \)^a $,  
satisfies~\cite{footnote1}
%{Einstein's summation 
%convention is adopted throughout this paper. }
%%%%%%%%%%%%%%%%%%%%%%%%%%%%%%%%%%%%%%
%%%%%%%%%%%%%%%%%%%%%%%%%%%%%%%%%%%%%%
\Beq
\xi^b \nb_b \xi^a =0\ \ {\rm with}\ \    \xi^a  \xi_a  = -1\ \ .
\label{eq:geodesic}
\Eeq
%%%%%%%%%%%%%%%%%%%%%%%%%%%%%%%%%%%%%%
%%%%%%%%%%%%%%%%%%%%%%%%%%%%%%%%%%%%%%

Let us take  a {\it timelike geodesic congruence} 
$\C$ over an open set $\Om$  in  $\M$; it means that  
we consider  some family $\C$ of timelike geodesics among which there is one and 
only one geodesic passing through any given point $p$ in $\Om$.  

We note that, for a given  $\Om$,  one can choose continuously infinite number of 
 timelike geodesic congruences over $\Om$ as is easily seen for the case of the Minkowski spacetime for $(\M, g)$. 

For a fixed timelike geodesic congruence $\C$,  
let  $\{ \g_s (\t) \}_{s \in (s_1, s_2)} $ be a smooth 1-parameter family of geodesics in $\C$:  Then  
$\{ \g_s (\t) \}_{s \in (s_1, s_2)} $ forms a smooth 
2-dimensional submanifold embedded in $\Om$
 with $\{ \t, s \}$ providing  a system 
 of smooth coordinates on it~\cite{footnote2}.  
By suitable parameterizations w.r.t. (with respect to)  
$s$ and suitable synchronizations 
w.r.t. $\t$ (i.e. choosing suitably  the  $\t=0$ point on  each member of the family), 
one can assume that the tangent vector $\xi_{s}^a (\t)$ of $\g_s (\t)$  and 
the {\it deviation vector} ${\eta_{s}}^a (\t)$, defined by 
${\eta_{s}}^a (\t) \equiv \(\del/\del s \)^a $, are orthogonal to each other 
for a fixed $s$. 
One can thus assume that 
geodesics in the 1-parameter family 
$\{ \g_s (\t) \}_{s \in (s_1, s_2)} $
 are so synchronized that all the $\t=const$ curves 
drawn within $\{ \g_s (\t) \}_{s \in (s_1, s_2)} $ are  orthogonal to 
the geodesics themselves.
%{For notational convenience, we shall assume 
%below that  the intervals such as  $(s_1, s_2)$ here and  $\Lambda$ in 
%Sec.{\ref{subsec:1-parameter}} contain $0$ whenever necessary.}. 

The deviation vector ${\eta}^a (\t)$ along a fixed geodesic  $\g$  is  measuring 
the deviation of the nearby geodesics among the family from 
being parallel to $\g$.

Here we  note that, for a fixed geodesic in the family, the pseudo-norm  of $\eta^a$,   $\eta^a  \eta_a$,  is not   constant along the geodesic  in general, 
contrary to the case of $\xi^a$ (Eq.(\ref{eq:geodesic})). This fact is important since 
it allows us to describe the convergence properties of the  geodesic congruence 
in terms of $\eta^a$ (see the arguments after Eq.(\ref{eq:B-decompose})).

\subsection{\label{subsec:expansion}Expansion $\th$, shear $\s_{ab}$ and 
twist $\om_{ab}$}

Let $\g$ be some timelike geodesic  and $\xi^a$ be its tangent vector as before. 
Then the symmetric covariant tensor $h_{ab}$ defined by  
%%%%%%%%%%%%%%%%%%%%%%%%%%%%%%%%%
%%%%%%%%%%%%%%%%%%%%%%%%%%%%%%%%%
\Beq
h_{ab} \equiv g_{ab} + \xi_a \xi_b  \ \ ,
\label{eq:hab}
\Eeq
%%%%%%%%%%%%%%%%%%%%%%%%%%%%%%%%%
%%%%%%%%%%%%%%%%%%%%%%%%%%%%%%%%%
satisfies  $h_{ab} \xi^b =0$ so that  it is regarded as 
a {\it spatial} tensor in the following sense. 

Let $T_p \M$ be the tangent space at a point $p$ in $\g$. 
Let ${T_p} \M^\perp$ be the $(n-1)$-dimensional subspace 
of $T_p \M$ consisting of  all the vectors orthogonal to $\xi^a$. 
Then ${T_p} \M^\perp$ represents  
all the vectors in the   spatial directions  for an observer moving along $\g$, so that 
vectors  and tensors constructed by ${T_p} \M^\perp$ are regarded 
as {\it spatial} objects for the observer. 
In particular,   when $h_{ab}$ in Eq.(\ref{eq:hab}) 
 is regarded  as  restricted  on ${T_p} \M^\perp$, 
the former  plays the role of the spatial, positive definite induced metric on the latter.

Furthermore, 
the $(1,1)$-tensor ${h_a}^b$,  obtained by raising one index of  $h_{ab}$ 
by  $g^{ab}$, 
satisfies ${h_a}^b {h_b}^c = {\d_a}^c$, playing the role of 
the projection operator 
%%%%%%%%%%%%%%%%%%%%%%%%%%%%%%%%%
%%%%%%%%%%%%%%%%%%%%%%%%%%%%%%%%
\Beq
h: T_p \M \lra  {T_p} \M^\perp \ \ .  
\label{eq:projection}
\Eeq
%%%%%%%%%%%%%%%%%%%%%%%%%%%%%%%%%%
%%%%%%%%%%%%%%%%%%%%%%%%%%%%%%%%%%%

Now, for a given 1-parameter family $\{\g_s\}_{s \in (s_1, s_2)}$ in a  timelike geodesic congruence $\C$,  we can estimate 
$\ddt \eta^a \equiv \nb_\xi \eta^a $, 
 the rate of change in the deviation vector $\eta^a$ 
 along a geodesic $\g$ $(\equiv \g_{s=0})$ in $\{\g_s\}_{s \in (s_1, s_2)}$.  
One can easily get~\cite{Wald} 
%%%%%%%%%%%%%%%%%%%%%%%%%%%%%%%%%%%%%
%%%%%%%%%%%%%%%%%%%%%%%%%%%%%%%%%%%%% 
\Beq
\ddt \eta^a ={B^a}_b \eta^b \ \ ,
\label{eq:deviation}
\Eeq
%%%%%%%%%%%%%%%%%%%%%%%%%%%%%%%%%%%%%
%%%%%%%%%%%%%%%%%%%%%%%%%%%%%%%%%%%%%%
where we have introduced
%%%%%%%%%%%%%%%%%%%%%%%%%%%%%%%%%%%%%%%
%%%%%%%%%%%%%%%%%%%%%%%%%%%%%%%%%%%%%%%
\Beq
B_{ab}\equiv  \nb_b \xi_a \ \ .
\label{eq:Bab}
\Eeq
%%%%%%%%%%%%%%%%%%%%%%%%%%%%%%%%%%%%%%%
%%%%%%%%%%%%%%%%%%%%%%%%%%%%%%%%%%%%%%%
To get Eq.(\ref{eq:deviation}), 
  we have used  $[\xi , \eta]^a \equiv  \nb_\xi \eta^a - \nb_\eta \xi^a =0 $, 
which follows due to 
the fact that  $\xi^a$ and  $\eta^a$ are vectors associated with 
 the coordinate functions $\t$ and $s$, respectively. 
Here we should note the  positions of indices $a$ and  $b$ 
in the definition of $B_{ab}$,  describing $\ddt \eta^a$ as 
a linear transformation in the conventional form as the R.H.S. (right-hand side) of 
 Eq.(\ref{eq:deviation}).     
We also note that, even though we have chosen the 1-parameter family 
$\{\g_s\}_{s \in (s_1, s_2)}$ in $\C$ for introducing $B_{ab}$, 
the latter describes the deformation properties of  $\C$ as a whole  
independent of the choice of the 1-parameter family  
$\{\g_s\}_{s \in (s_1, s_2)}$.

It is important  that $B_{ab}$ is a spatial tensor, satisfying
\[
B_{ab}\xi^b =0 \ \ , \ \  \xi^a B_{ab}=0 \ \ ,
\]
as is  shown with the help  of  Eq.(\ref{eq:geodesic}). 

The tensor $B_{ab}$ plays the central role in  our analysis below. 
It is convenient to decompose $B_{ab}$ into the following form; 
%%%%%%%%%%%%%%%%%%%%%%%%%%%%%%%
%%%%%%%%%%%%%%%%%%%%%%%%%%%%%%%
\Beq
B_{ab} = \frac{1}{n-1} \th h_{ab} + \s_{ab} + \om_{ab}\ \ .  
\label{eq:B-decompose}
\Eeq
%%%%%%%%%%%%%%%%%%%%%%%%%%%%%%%
%%%%%%%%%%%%%%%%%%%%%%%%%%%%%%%
Here the first term on the R.H.S. is the trace-part of $B_{ab}$ with 
 $\th\equiv  {B_a}^a$;  the second and the third terms are 
both  trace-free   with the former and the latter being 
 symmetric  and  anti-symmetric in the indices, respectively.     
Accordingly it follows that 
$\s_{ba}= \s_{ab}$ with ${\s_a}^a =0$ and $\om_{ba}=-\om_{ab}$. 

It is conventional to call  $\th$, $\s_{ab}$ and $\om_{ab}$  the {\it expansion}, the {\it shear} and the {\it twist}, respectively.  

Due to Eq.(\ref{eq:deviation}),  which is in  the standard form describing 
deformations in the linear transformations, one can convince oneself of 
the interpretations for $\th$, $\s_{ab}$ and $\om_{ab}$. 
For a given timelike geodesic congruence $\C$ (synchronized  as discussed in Sec.\ref{subsec:congruence}),  the set ${\mathcal{S}}_{\t}$  
of  the points   with  the affine-parameter  $\t$
 forms an $(n-1)$-dimensional section of $\C$. 
When comparing ${\mathcal{S}}_{\t}$ and ${\mathcal{S}}_{\t+\D \t}$, then, 
$\th$, $\s_{ab}$ and $\om_{ab}$ are interpreted as the rate of 
 the dilation of the  $(n-1)$-volume,  the shape-change and 
the rotation, respectively, averaged during the time-interval $\D \t$.

By the Frobenius's theorem, $\om_{ab}=0$ holds  {\sl iff}  
the timelike geodesic congruence $\C$ is {\it hypersurface orthogonal}, i.e. 
the whole of  $\C$ can be foliated by smooth $(n-1)$-dimensional 
spatial sections orthogonal to every  timelike geodesic contained in 
$\C$~\cite{Wald}.  
In such cases, $B_{ab}$ reduces to  
a spatial symmetric tensor,   coinciding with the extrinsic curvature $K_{ab}$ 
for  the smooth spatial sections orthogonal to all timelike geodesics in $\C$.   
At the same time, 
$\om_{ab}=0$ holds {\sl iff} the spatial covariant derivative $\bmD_a$
  is torsion-free (see Eq.(\ref{eq:bmD-Torsion})  and arguments following it 
 in {\it Appendix} \ref{app:formulas}). 
Here  $\bmD_a$ is the  spatial derivative operator on $T \M^\perp$ 
 induced from   $\nb_a$ through the projection $h$  (Eq.(\ref{eq:projection})), 
defined as 
%%%%%%%%%%%%%%%%%%%%%%%%%%%%%%%%%%%
%%%%%%%%%%%%%%%%%%%%%%%%%%%%%%%%%%
\Beq
\bmD_a T^{b\cdots}_{c\cdots} 
\equiv 
{h_a}^p {h^b}_q \cdots {h_c}^r \cdots   \nb_p T^{q\cdots}_{r\cdots}
\label{eq:bmD}
\Eeq
%%%%%%%%%%%%%%%%%%%%%%%%%%%%%%%%%%%%%%
%%%%%%%%%%%%%%%%%%%%%%%%%%%%%%%%%%%%%
for any tensor $T^{a\cdots}_{b\cdots}$.

\subsection{\label{subsec:notations}Notations}
At this point, let us summarize  the notations employed below.
%\vskip .5cm
\BD
\item[(i)]
We often adopt  the standard notations for the products among 
matrices and vectors when they  save indices. For instance, 
 second-rank tensors ${A^a}_b$ and  ${B^a}_b$ can be regarded as  matrices.  
Then 
\[
{(AB)^a}_b \equiv {A^a}_c {B^c}_b\ \ , \ \ 
(Au)^a \equiv {A^a}_b u^b\ \ .
\] 
 We also allow the  remaining indices to be raised or lowered freely as usual tensorial 
indices.   For instance, 
\[
(AB)_{ab} \equiv A_{ac} {B^c}_b \ \ .
\] 
\item[(ii)]
For vectors $u^a$ and $v^a$, we define 
\[
u \cdot v \equiv u_a v^a\ \ .
\]
 Similarly, 
for  tensors $\a_{ab}$ and $\b_{ab}$, we define 
(note the index-positions)
\[
(\a \cdot \b)_{ab}\equiv \a_{ac} {\b_b\ }^c \ \ , 
\]   
and 
\[
\a \cdot \b \equiv {(\a \cdot \b)_{a}}^a \equiv \a_{ab} \b^{ab}\ \ .
\] 
 Note the difference between 
$(\a \cdot \b)_{ab}$ defined here and 
$(\a \b)_{ab} \equiv {\a_a}^c \b_{cb}$ defined in {\bf (i)}.  
\item[(iii)]
Overlines (underlines) attached to more than two   indices indicate symmetrization 
(anti-symmetrization) of  these indices.  
For instance, 
%%%%%%%%%%%%%%%%%%%%%%%%%%%%%%%%%
\Beaa
&& \a_{\ol{a}\ol{b}}\equiv \frac{1}{2} (\a_{ab}+\a_{ba}) \ \ ,  \ \ 
 \a_{\ol{a} b  \ol{c} d}\equiv \frac{1}{2} (\a_{abcd}+\a_{cbad}) \ \ , \\
&& \b_{\ul{a}\ul{b}}\equiv \frac{1}{2} (\b_{ab}-\b_{ba}) \ \ , \ \ 
 \b_{\ul{a} b  \ul{c} d}\equiv \frac{1}{2} (\b_{abcd}-\b_{cbad})\ \ .
\Eeaa
(Anti-)symmetrizations are operated prior to  contractions when they appear simultaneously. For instance,
\[
g^{ac}(\g B)_{\ul{c}\ul{b}} \equiv \half g^{ac} (\g_{cd} {B^d}_b - \g_{bd} {B^d}_c  )\ \ .
\]
%%%%%%%%%%%%%%%%%%%%%%%%%%%%%%%%%%%
\item[(iv)]
Any index $a$ is replaced by a dot ( `` $ {}^\Dot $ " ) when it is contracted with 
a geodesic tangent vector $\xi^a$.
For instance, 
\[
u_{a^\Dot}\equiv u_{ab} \xi^b \ \ , \ \  
v^{a^\Dot} \equiv v^{ab} \xi_b \ \ , \ \  
\gddot \equiv \g_{ab} \xi^a \xi^b\ \ .
\] 
When no confusions are caused, these dots may often be omitted.  
For instance, $\g^{\, a}$ can be used as a shorthand  for  $\g^{\, a^\Dot}$ 
provided that it is obvious in the context. 
\item[(v)]
A second rank {\it spatial} tensor with the tilde-symbol  (``\ \ $\widetilde{}\ \ $")  
  indicates  its  trace-free part.  For instance,  
when   $\a_{ab}$ is a
 spatial tensor,   
\[
{\widetilde{\a}}_{ab}
\equiv \a_{ab} - \frac{1}{n-1} {\a_c}^c h_{ab}\ \ .
\]
  Furthermore, 
 $(\a \cdot \a)_{ab}$ is also a spatial tensor.  Then,    
\[
\widetilde{(\a \cdot \a)}{}_{ab}
\equiv (\a \cdot \a)_{ab} - \frac{1}{n-1} (\a \cdot \a) h_{ab}\ \ .
\] 
\item[(vi)]
The projection map Eq.(\ref{eq:projection}) induces 
maps from  tensors to their corresponding spatial tensors. 
These maps are denoted by  the ``underline" symbol attached to 
the main letter representing the mapped object.
For instance, 
\[
{\ul{v}\, }_a \equiv {h_a}^b v_b\ \ , \ \ 
{\ul{\g}\, }_{ab}\equiv {h_a}^c {h_b}^d \g_{cd}\ \ .
\]     
We also note that, for instance,  $\widetilde{{\ul{R}}}_{\, ab}$ 
unambiguously indicates the trace-free part of the spatial projection 
of $R_{ab}$, noting that 
 ``\ \  $\widetilde{}$ \ " is meaningful only for the {\it spatial} objects ({\bf (v)}). 
\ED

\subsection{\label{subsec:Raychaudhuri}Raychaudhuri-type equations}

Now, to estimate the rate of change of  $B_{ab}$ along a geodesic, 
one takes its derivative w.r.t. $\t$. 
Noting that $\ddt=\nb_{\xi}$ and taking into account 
Eq.(\ref{eq:geodesic}), 
it is then straightforward to get
%%%%%%%%%%%%%%%%%%%%%%%%%%%%%%%%%%%%%%
%%%%%%%%%%%%%%%%%%%%%%%%%%%%%%%%%%%%%%
\Beq
\frac{d}{d \t} B_{ab}   = -(BB)_{ab} - R_{a^\Dot  b^\Dot} \ \ , 
\label{eq:dBdt}
\Eeq 
%%%%%%%%%%%%%%%%%%%%%%%%%%%%%%%%%%%%%%
%%%%%%%%%%%%%%%%%%%%%%%%%%%%%%%%%%%%%%
where we adopt the standard definitions for the  curvatures~\cite{Wald}, 
%%%%%%%%%%%%%%%%%%%%%%%%%%%%%%%%%%%%%
%%%%%%%%%%%%%%%%%%%%%%%%%%%%%%%%%%%%%
\Bea 
&& (\nb_a \nb_b - \nb_b \nb_a) u_c = {R_{abc}}^d u_d  \ \ {\rm for }\ \ 
\forall u^a, \nonumber \\
&& R_{ab} \equiv {R_{acb}}^c \ \ ,  \ \  R \equiv {R_a}^a\ \ .
\label{eq:Riemann}
\Eea
%%%%%%%%%%%%%%%%%%%%%%%%%%%%%%%%%%%%%
%%%%%%%%%%%%%%%%%%%%%%%%%%%%%%%%%%%%%
We note that $\ddt B_{ab}$ 
is a spatial tensor as is seen by Eq.(\ref{eq:dBdt}). In general, 
$\ddt A_{abc \cdots }$ is spatial when $A_{abc \cdots }$ is spatial, which 
is easily shown as 
%%%%%%%%%%%%%%%%%%%%%%%%%%%%%%%%%%%%%
%%%%%%%%%%%%%%%%%%%%%%%%%%%%%%%%%%%%%
\[
\xi^a \frac{d }{d \t} A_{abc \cdots }
= \frac{d }{d \t} \(\xi^a A_{abc \cdots } \)=0\ \  .
\]
%%%%%%%%%%%%%%%%%%%%%%%%%%%%%%%%%%%%%
%%%%%%%%%%%%%%%%%%%%%%%%%%%%%%%%%%%%%

By decomposing  the both sides of 
Eq.(\ref{eq:dBdt}) into the trace-part and the trace-free part, 
and by decomposing the latter into 
the symmetric and  anti-symmetric parts, 
we finally obtain
%%%%%%%%%%%%%%%%%%%%%%%%%%%%%
%%%%%%%%%%%%%%%%%%%%%%%%%%%%%
\Bea
&& \frac{d \th}{d\t}  = -\frac{1}{n-1}\ \th^2 - \s \cdot \s +\om \cdot \om 
- R_{ {}^\Dot {}^\Dot }\ \ ,  \nonumber \\
&& \frac{d \s_{ab}}{d\t}  
 = - \frac{2}{n-1}\  \th\  \s_{ab} - \widetilde{(\s \cdot \s)}_{ab} 
     + \widetilde{(\om  \cdot \om)}_{ab}  -\widetilde{R}_{a^\Dot  b^\Dot} \ \ , 
\nonumber  \\
&& \frac{d \om_{ab}}{d\t}  = - \frac{2}{n-1}\  \th \  \om_{ab} 
-2 (\om \cdot \s)_{\ul{a}\ul{b}} \ \ , \ \  
\label{eq:Raych}
\Eea
%\vskip .5cm
%%%%%%%%%%%%%%%%%%%%%%%%%%%%%%%%%%%%%%%%%%%%%%%%%%%%%%%%%%%%%%%%%
where $\widetilde{R}_{a^\Dot  b^\Dot}$ is  the trace-free part of 
$R_{a^\Dot  b^\Dot }$ regarded as a second-rank spatial tensor (see {\bf (iv)} and 
{\bf (v)} in Sec.\ref{subsec:notations}).  
Furthermore,  it is  easy to see that  
$\widetilde{R}_{a^\Dot  b^\Dot}$ is  expressed as
%%%%%%%%%%%%%%%%%%%%%%%%%%%%%%%%%
%%%%%%%%%%%%%%%%%%%%%%%%%%%%%%%%%
\Beq
\widetilde{R}_{a^\Dot  b^\Dot}
=C_{a^\Dot  b^\Dot} 
                             - \frac{1}{n-2}\widetilde{{\ul{R}}}_{\, ab}\ \ ,
\label{eq:Weyl}
\Eeq
%%%%%%%%%%%%%%%%%%%%%%%%%%%%%%%%%
%%%%%%%%%%%%%%%%%%%%%%%%%%%%%%%%%%%
where $C_{abcd}$ is the Weyl tensor.  

The first equation in Eq.(\ref{eq:Raych}) is often called
the {\it Raychaudhuri equation}~\cite{Raych1,Raych2,KarSengupta}, 
which plays the central role for proving the singularity theorems. 
For brevity let us  call the three equations in Eq.(\ref{eq:Raych}) 
the {\it Raychaudhuri-type equations}. 

\section{\label{sec:Deviation}Deviations  of  Raychaudhuri-type equations due to 
geometrical variations}
\subsection{\label{subsec:1-parameter}One-parameter family of geometries on $\M$}
Let us consider a 1-parameter family of $n$-dimensional spacetime geometries 
on $\M$,  
$\{(\M, g^{({\l})})\}_{\l \in \Lambda}$.
Let $\nb^{(\l)}$ be the covariant derivative compatible with $g_{ab}^{(\l)}$. 

For   some  quantity  $A(\l)$  regarded as a smooth function of $\l$,  
let  $\stackrel{\circ}{A}$ indicate 
the derivative of $A(\l)$ w.r.t. $\l$ at $\l=0$ and call 
the {\it $\circ$-derivative} of  $A(\l)$ for brevity.

For notational simplicity, we shall  write just $A$ for $A(0)$, such as 
 $\nb$ and $g_{ab}$ for $\nb^{(0)}$ and $g_{ab}^{(0)}$, 
respectively. 

We here introduce two quantities in connection with the $\circ$-derivative  
for later analysis,
%%%%%%%%%%%%%%%%%%%%%%%%%%%%%%%%%
%%%%%%%%%%%%%%%%%%%%%%%%%%%%%%%%%%%%
\Beq
\g_{ab} \equiv \stackrel{\circ}{g}_{ab}  \ \ ,  \ \ 
\chi^a \equiv \stackrel{\circ}{\xi^a}\ \ .
\label{eq:g-dot}
\Eeq
%%%%%%%%%%%%%%%%%%%%%%%%%%%%%%%%%%%%%
%%%%%%%%%%%%%%%%%%%%%%%%%%%%%%%%%%%%%
The former represents the geometrical variation  while the latter describes 
the shift  of a  geodesic due to the geometrical variation.

By taking the $\circ$-derivative on both sides of $g_{ab} g^{bc}= {\d_a}^c$ and 
$\xi_a = g_{ab} \xi^b$, 
we also get 
%%%%%%%%%%%%%%%%%%%%%%%%%%%%%%%%%%%%%
%%%%%%%%%%%%%%%%%%%%%%%%%%%%%%%%%%%%%
\Beq
\stackrel{\circ}{g}{}^{ab} = -\g^{ab}\ \ , \ \ 
\stackrel{\circ}{\xi}{}_a = \chi_a + \g_a \ \ . 
\label{eq:g-dot2}
\Eeq
%%%%%%%%%%%%%%%%%%%%%%%%%%%%%%%%%%%%%
%%%%%%%%%%%%%%%%%%%%%%%%%%%%%%%%%%%%%
Taking the $\circ$-derivative on both sides of $\xi^a \xi_a =-1$, we get 
a useful formula
%%%%%%%%%%%%%%%%%%%%%%%%%%%%%%%%%%%%%
%%%%%%%%%%%%%%%%%%%%%%%%%%%%%%%%%%%%%
\Beq
\xi \cdot \chi = -\half \gddot\ \ ,
\label{eq:xi-chi}
\Eeq
%%%%%%%%%%%%%%%%%%%%%%%%%%%%%%%%%%%%%
%%%%%%%%%%%%%%%%%%%%%%%%%%%%%%%%%%%%%
where Eq.(\ref{eq:g-dot}) and Eq.(\ref{eq:g-dot2}) have been used.

Now, according to  general properties of covariant derivatives~\cite{Wald}, 
for any vector $w_a$, the difference between  
$\nb_a w_b$ and $\nb_a^{(\l)}  w_b$ should be represented as
${\C^c}_{ab}(\l) w_c$  with 
${\C^c}_{ab}(\l)$ being some function of $\l$, 
%%%%%%%%%%%%%%%%%%%%%%%%%%%%
%%%%%%%%%%%%%%%%%%%%%%%%%%%%
\Beq
\nb_a^{(\l)} w_b =\nb_a w_b - {\C^c}_{ab}(\l) w_c\ \ .
\label{eq:DDC}
\Eeq
%%%%%%%%%%%%%%%%%%%%%%%%%%%%%%
%%%%%%%%%%%%%%%%%%%%%%%%%%%%%%
Applying Eq.(\ref{eq:DDC}) to the identity  $\nb^{(\l)}_a g_{bc}^{(\l)}=0$ and 
following the standard procedure, 
 ${\C^c}_{ab}(\l)$ can be expressed as 
%%%%%%%%%%%%%%%%%%%%%%%%%%%%%%%%%
%%%%%%%%%%%%%%%%%%%%%%%%%%%%%%%%%
\Beq
{\C^c}_{ab}(\l) 
=\frac{1}{2}g^{cd\, (\l)} \left( \nb_a g_{bd}^{(\l)} +  \nb_b g_{ad}^{(\l)}
- \nb_d g_{ab}^{(\l)}  \right)\ \ .
\label{eq:Cabc}
\Eeq
%%%%%%%%%%%%%%%%%%%%%%%%%%%%%%%%%
%%%%%%%%%%%%%%%%%%%%%%%%%%%%%%%%%
Here $\nb_a$ is the covariant derivative compatible to $g_{ab}$ so that it 
 is independent of $\l$.

Taking the $\circ$-derivative on both size of Eq.(\ref{eq:Cabc}), then, it follows that 
%%%%%%%%%%%%%%%%%%%%%%%%%%%%%%%%%
%%%%%%%%%%%%%%%%%%%%%%%%%%%%%%%%%
\Beq
{\stackrel{\circ}{\C^c}}_{ab}
=\frac{1}{2}g^{cd} \left( \nb_a \g_{bd} +  \nb_b \g_{ad}
- \nb_d \g_{ab}  \right)\ \ . 
\label{eq:Cdot}
\Eeq
%%%%%%%%%%%%%%%%%%%%%%%%%%%%%%%%%
%%%%%%%%%%%%%%%%%%%%%%%%%%%%%%%%%
From Eq.(\ref{eq:Cdot}), we get a useful formula
%%%%%%%%%%%%%%%%%%%%%%%%%%%%%%%%%
%%%%%%%%%%%%%%%%%%%%%%%%%%%%%%%%%
\Beq
{\stackrel{\circ}{\C^c}}_{ab}\xi^a \xi^b
= \ddt {\g\, }^c  - \frac{1}{2} (\nb^c \g_{ab}) \xi^a \xi^b \ \ . 
\label{eq:Cdotxixi}
\Eeq
%%%%%%%%%%%%%%%%%%%%%%%%%%%%%%%%%
%%%%%%%%%%%%%%%%%%%%%%%%%%%%%%%%%
\subsection{\label{subsec:geodesic_shift} Key equation for 
geodesic shift  due to geometrical variations}
%%%%%%%%%%%%%%%%%%%%%%%%%%%%%%%%%
%%%%%%%%%%%%%%%%%%%%%%%%%%%%%%%%%

When a spacetime geometry is varied, geodesics should be shifted accordingly. 
We  now derive a key equation describing such a geodesic 
shift  caused by geometry variations. 

Consider a  geodesic  $\g^{(\l)}$ in the spacetime $(\M, g^{(\l)})$. Its tangent vector, 
$\xi^a (\l)$, should satisfy
%%%%%%%%%%%%%%%%%%%%%%%%%%%%%%%%%%%%%%
%%%%%%%%%%%%%%%%%%%%%%%%%%%%%%%%%%%%%%%
\Beq
\xi^b (\l) \nb_b^{(\l)} \xi^a (\l)=0\ \ .
\label{eq:Lgeodesic}
\Eeq
%%%%%%%%%%%%%%%%%%%%%%%%%%%%%%%%%%%%%%%%
%%%%%%%%%%%%%%%%%%%%%%%%%%%%%%%%%%%%%%%%
Taking the $\circ$-derivative on both sides of Eq.(\ref{eq:Lgeodesic}), we get
%%%%%%%%%%%%%%%%%%%%%%%%%%%%%%%%%%%%%%
%%%%%%%%%%%%%%%%%%%%%%%%%%%%%%%%%%%%%%%
\[
\chi^b {B^a}_b 
+ \xi^b (\nb_b \chi^a +  {\stackrel{\circ}{\C^a}}_{bc} \xi^c ) =0\ \ , 
\]
%%%%%%%%%%%%%%%%%%%%%%%%%%%%%%%%%%%%%%%%
%%%%%%%%%%%%%%%%%%%%%%%%%%%%%%%%%%%%%%%%
which along with  Eq.(\ref{eq:Cdotxixi}) yields 
%%%%%%%%%%%%%%%%%%%%%%%%%%%%%%%%%%%%%%
%%%%%%%%%%%%%%%%%%%%%%%%%%%%%%%%%%%%%%%
\Beq
\ddt \chi^a + {B^a}_b \chi^b= 
-\ddt {\g\, }^a  + \frac{1}{2} (\nb^a \g_{bc}) \xi^b \xi^c\ \ .
\label{eq:keyeq_chi}
\Eeq
%%%%%%%%%%%%%%%%%%%%%%%%%%%%%%%%%%%%%%%%
%%%%%%%%%%%%%%%%%%%%%%%%%%%%%%%%%%%%%%%%
What we are concerned with is, however,  the deviation of the geodesic  
from  the original  geodesic $\g$  caused by  geometry variations ($\g_{ab}$), 
 so that   it is more appropriate to use ${\chi_\perp}^a$, 
the component of $\chi^a$ orthogonal to 
$\xi^a$, rather than $\chi^a$ itself. 
Indeed, it turns out that 
the formulas below become much simpler and more transparent in meaning 
in terms of 
${\chi_\perp}^a$ ~\cite{footnote3}.
%{Here the concepts of ``orthogonal" and 
%``parallel" are naturally judged in the original geometry $(\M , g)$.}.  
 On the other hand, the component ${\chi_\parallel}^a$ which is  parallel to $\xi^a$ does not shift $\g$,  so that  it may be regarded as the ``gauge-freedom" in the present description. In effect, we  fix the gauge-freedom by  ${\chi_\parallel}^a \equiv 0$.

Let us then introduce
%%%%%%%%%%%%%%%%%%%%%%%%%%%%%%%%%%%
%%%%%%%%%%%%%%%%%%%%%%%%%%%%%%%%%%
\Beq
\nu^a \equiv {\chi_\perp}^a={\ul{\chi}}^a \ \  , 
\label{eq:nu1}
\Eeq
%%%%%%%%%%%%%%%%%%%%%%%%%%%%%%%%%%
%%%%%%%%%%%%%%%%%%%%%%%%%%%%%%%%%%
which  satisfies with the help of Eq.(\ref{eq:xi-chi})
%%%%%%%%%%%%%%%%%%%%%%%%%%%%%%%%%%%
%%%%%%%%%%%%%%%%%%%%%%%%%%%%%%%%%%
\Bea
&& \nu^a = \chi^a - \frac{1}{2}\gddot \xi^a\ \ , \nonumber  \\
&& \xi \cdot \nu =\xi \cdot \frac{d\nu}{d\t} = 0\ \ .  
\label{eq:nu2}
\Eea
%%%%%%%%%%%%%%%%%%%%%%%%%%%%%%%%%%%%%%
%%%%%%%%%%%%%%%%%%%%%%%%%%%%%%%%%%%%%%%

Considering Eq.(\ref{eq:g-dot}) along with Eq.(\ref{eq:nu2}), we note
%%%%%%%%%%%%%%%%%%%%%%%%%%%%%%%%%%%
%%%%%%%%%%%%%%%%%%%%%%%%%%%%%%%%%%
\Bea
&&   {\stackrel{\circ}{\xi}}{}^a
         = \nu^a + \frac{1}{2}\gddot \xi^a\ \ , \nonumber  \\
&&   {\stackrel{\circ}{\xi}}{}_a
         = \nu_a + \g_a+ \frac{1}{2}\gddot \xi_a\ \ . 
\label{eq:xidot}
\Eea
%%%%%%%%%%%%%%%%%%%%%%%%%%%%%%%%%%%%%%
%%%%%%%%%%%%%%%%%%%%%%%%%%%%%%%%%%%%%%%

Now Eq.(\ref{eq:keyeq_chi}) can be represented in terms of $\nu^a$ 
by means of  Eq.(\ref{eq:nu2}); 
%%%%%%%%%%%%%%%%%%%%%%%%%%%%%%%%%%%%%%
%%%%%%%%%%%%%%%%%%%%%%%%%%%%%%%%%%%%%%%
\Beq
\ddt \nu^a + {B^a}_b \nu^b= 
\frac{1}{2} \bmD^a \gddot - \ddt{\ul{\g}}^a\ \ ,
\label{eq:keyeq}
\Eeq
%%%%%%%%%%%%%%%%%%%%%%%%%%%%%%%%%%%%%%%%
%%%%%%%%%%%%%%%%%%%%%%%%%%%%%%%%%%%%%%%%
where  $\bmD_a$ is the  spatial derivative operator  
 induced from   $\nb_a$  (Eq.(\ref{eq:bmD})).

It is obvious that Eq.(\ref{eq:keyeq}), the relation solely among   spatial quantities, 
  is much more desirable than Eq.(\ref{eq:keyeq_chi}). 

Equation (\ref{eq:keyeq}) is our key equation, telling us the 
linear response of the geodesic shift $\nu^a$  to the geometry variation $\g_{ab}$.

Equation (\ref{eq:keyeq}) is also expressed as
%%%%%%%%%%%%%%%%%%%%%%%%%%%%%%%%%%%%%%
%%%%%%%%%%%%%%%%%%%%%%%%%%%%%%%%%%%%%%
\Beq
{\mathcal{L}^a}_b \nu^b = 
\frac{1}{2} \bmD^a \gddot - \ddt{\ul{\g}}^a\ \ , 
\label{eq:Lop}
\Eeq
%%%%%%%%%%%%%%%%%%%%%%%%%%%%%%%%%%%%%%
%%%%%%%%%%%%%%%%%%%%%%%%%%%%%%%%%%%%%%
with
%%%%%%%%%%%%%%%%%%%%%%%%%%%%%%%%%%%%%%%%%
%%%%%%%%%%%%%%%%%%%%%%%%%%%%%%%%%%%%%%%
\Beq
{\mathcal{L}^a}_b \equiv \ddt {\d^a}_b + {B^a}_b\ \ .
\label{eq:Leq}
\Eeq
%%%%%%%%%%%%%%%%%%%%%%%%%%%%%%%%%%%%%%%
%%%%%%%%%%%%%%%%%%%%%%%%%%%%%%%%%%%%%%%
 Since $\nu^a$ (rather than $\chi^a$) contains the pure geometrical information 
on the geodesic shift, 
it should hold $\nu^a \equiv 0 \iff \g_{ab} \equiv 0$, so that 
${\mathcal{L}^a}_b$ should be invertible. Thus one can formally solve 
Eq.(\ref{eq:Lop}) as 
%%%%%%%%%%%%%%%%%%%%%%%%%%%%%%%%%%%%
%%%%%%%%%%%%%%%%%%%%%%%%%%%%%%%%%%%%%
\Bea
&&  \nu^a =  {\(\mathcal{L}^{-1}\)^a}_b 
{
\( \frac{1}{2} \bmD \gddot - \ddt{\ul{\g}} \)
}^b \nonumber \\
&& \quad \ \ = \Biggl[
\left(\ddt\bm{1} +  \bm{B} \right)^{-1}
\(\frac{1}{2} \bmD  \gddot - \ddt{\ul{\g}} \)
\Biggr]^a \ \ ,
\label{eq:nusolve}
\Eea
%%%%%%%%%%%%%%%%%%%%%%%%%%%%%%%%%%%%%
%%%%%%%%%%%%%%%%%%%%%%%%%%%%%%%%%%%%% 
which formally expresses the geodesic shift $\nu^a$ in terms of 
the geometry variation $\g_{ab}$.

%%%%%%%%%%%%%%%%%%%%%%%%%%%%%%%%%
%%%%%%%%%%%%%%%%%%%%%%%%%%%%%%%%%
\section{\label{sec:Convergence_shift} Changes in  geodesic
 convergence properties due to geometry variations}
\subsection{\label{subsection:B-dot} Deviations of  $\th$, $\s_{ab}$ and $\om_{ab}$ 
due to geometry variations}

Now taking the $\circ$-derivative of   
%%%%%%%%%%%%%%%%%%%%%%%%%%%%%%%%%%%%
%%%%%%%%%%%%%%%%%%%%%%%%%%%%%%%%%%%%
\[
B_{ab}(\l) = \nb_{b\> (\l)} \xi_a (\l) = \nb_b \xi_a (\l)  - {\C^c}_{ab}(\l) \xi_c(\l)
\ \ ,
\]
%%%%%%%%%%%%%%%%%%%%%%%%%%%%%%%%%%%%
%%%%%%%%%%%%%%%%%%%%%%%%%%%%%%%%%%%%
we get after some straightforward calculations, 
%%%%%%%%%%%%%%%%%%%%%%%%%%%%%%%%%%%%
%%%%%%%%%%%%%%%%%%%%%%%%%%%%%%%%%%%%
\Bea
&&  {\stackrel{\circ}{B}}_{ab} 
                   = (\g B)_{\ol{a}\ol{b}}+ \frac{1}{2}\gddot B_{ba}
                     + \frac{1}{2} \ddt {\ul{\g}\, }_{ab} 
                                         +  \bmD_b \nu_a \nonumber \\ 
&&  \quad  + (B \nu)_a \xi_b + \xi_a (\nu B)_b 
          + \bmD_{\ul{b}}\, \ul{\g}\, {}_{\ul{a}}
          + \xi_{\ul{a}} (\ul{\g} B)_{\ul{b}}\ \ .
\label{eq:Bdot1}
\Eea
%%%%%%%%%%%%%%%%%%%%%%%%%%%%%%%%%%%%%%
%%%%%%%%%%%%%%%%%%%%%%%%%%%%%%%%%%%%%%
Here we note that, according to Sec.{\ref{subsec:notations}}, 
all the  terms are unambiguously  defined in shorthand notations. 
Namely,  $(\g B)_{\ol{a}\ol{b}} = \g_{\ol{a} c}{B^c}_{\ol{b}}$ , 
$(B \nu)_a= B_{ac} \nu^c$ ,    $(\nu B)_b=\nu^c B_{cb}$ , 
$\xi_{\ul{a}} (\ul{\g} B)_{\ul{b}}
=\xi_{\ul{a}} {\ul{\g}}{}_{c d} \xi^d {B^c}_{\ul{b}} $  and so forth. 
\vskip .2cm

Contrary to  the case of $\ddt B_{ab}$ (see Eq.(\ref{eq:dBdt})),  
${\stackrel{\circ}{B}}_{ab} $ is not spatial any more as is seen in 
the second line on the R.H.S. of Eq.(\ref{eq:Bdot1}). 
This result turns out to be a reasonable one, repeating the similar argument after 
Eq.(\ref{eq:Riemann}) along with $\stackrel{\circ}{\xi^a} \neq 0$. 

Similarly we also get
%%%%%%%%%%%%%%%%%%%%%%%%%%%%%%%%%%%%
%%%%%%%%%%%%%%%%%%%%%%%%%%%%%%%%%%%%
\Bea
&&  {{\stackrel{\circ}{B}{}^a}}_b
                   =-  g^{ac} (\g B)_{\ul{c}\ul{b}}
                        + \frac{1}{2}\gddot {B_b}^a  
                     + \frac{1}{2} \ddt {{\ul{\g}\, }^a}_b 
                                         +  \bmD_b \nu^a \nonumber \\ 
&&  \ \  + (B \nu)^a \xi_b + \xi^a (\nu B)_b 
          + g^{ac} \bigl( \bmD_{\ul{b}}\, \ul{\g}\, {}_{\ul{c}}
          + \xi_{\ul{c}} (\ul{\g} B)_{\ul{b}} \bigr)\ \ .
\label{eq:Bdot2}
\Eea
%%%%%%%%%%%%%%%%%%%%%%%%%%%%%%%%%%%%%%
%%%%%%%%%%%%%%%%%%%%%%%%%%%%%%%%%%%%%%
Here the difference between the first terms in Eq.(\ref{eq:Bdot1}) and 
Eq.(\ref{eq:Bdot2}) should be noted.  (Note the signs and 
the (anti-)symmetrization. ) 

By contracting  both sides of  Eq.(\ref{eq:Bdot1}) with $g^{ab}$ along 
with simple manipulations, we get the formula for $\stackrel{\circ}{\th}$ as
%%%%%%%%%%%%%%%%%%%%%%%%%%%%%%%%%%%%
%%%%%%%%%%%%%%%%%%%%%%%%%%%%%%%%%%%%
\Bea
\stackrel{\circ}{\th} = \half \gddot \th + \half \ddt \ul{\g} + \bmD \cdot \nu \ \ ,
\label{eq:th-dot}
\Eea
%%%%%%%%%%%%%%%%%%%%%%%%%%%%%%%%%%%%%%
%%%%%%%%%%%%%%%%%%%%%%%%%%%%%%%%%%%%%%
 where $\ul{\g}$ is the shorthand notation for $h^{ab} \ul{\g}{}_{ab}$, the trace 
of $\ul{\g}{}_{ab}$. 
Equation (\ref{eq:th-dot}) is one of our basic formulas describing 
changes in  geodesic convergence properties caused by geometry variations. 

To get similar basic formulas for $\stackrel{\circ}{\s}{}_{ab}$ and 
$\stackrel{\circ}{\om}{}_{ab}$, some more considerations are helpful.
Firstly  we note that 
%%%%%%%%%%%%%%%%%%%%%%%%%%%%%%%%%%%%%
%%%%%%%%%%%%%%%%%%%%%%%%%%%%%%%%%%%%%
\Bea
&& \stackrel{\circ}{\s}{}_{ab}-\stackrel{\circ}{\s}{}_{ba} 
= (\s_{ab} - \s_{ba})^\circ = 0 \ \ , \nonumber \\
&& \stackrel{\circ}{\om}{}_{ab}+\stackrel{\circ}{\om}{}_{ba} 
= (\om_{ab} + \om_{ba})^\circ = 0 \ \ ,  
\label{eq:s-om-symmetry}
\Eea
%%%%%%%%%%%%%%%%%%%%%%%%%%%%%%%%%%%%%
%%%%%%%%%%%%%%%%%%%%%%%%%%%%%%%%%%%%%
due to the linearity of the $\circ$-derivative. Thus, 
$\stackrel{\circ}{\s}{}_{ab}$ and $\stackrel{\circ}{\om}{}_{ab}$ are 
 symmetric and anti-symmetric  in the indices, respectively  just as 
$\s_{ab}$ and $\om_{ab}$. 

Now taking the $\circ$-derivative on both sides of  Eq.(\ref{eq:B-decompose}), 
we see  
%%%%%%%%%%%%%%%%%%%%%%%%%%%%%%%%%%%%
%%%%%%%%%%%%%%%%%%%%%%%%%%%%%%%%%%%%
\Beq
\stackrel{\circ}{\s}{}_{ab} + \stackrel{\circ}{\om}{}_{ab}   
=  \stackrel{\circ}{B}{}_{ab} 
 - \frac{1}{n-1} \stackrel{\circ}{\th} h_{ab}  - 
\frac{1}{n-1} \th \stackrel{\circ}{h}{}_{ab}
\ \ .
\label{eq:s-dot-om-dot}
\Eeq
%%%%%%%%%%%%%%%%%%%%%%%%%%%%%%%%%%%%
%%%%%%%%%%%%%%%%%%%%%%%%%%%%%%%%%%%%
Considering Eq.(\ref{eq:s-om-symmetry}), then, we can get the formulas 
for $\stackrel{\circ}{\s}{}_{ab}$ and $\stackrel{\circ}{\om}{}_{ab}$
by decomposing the R.H.S. of Eq.(\ref{eq:s-dot-om-dot}) into 
the symmetric part and the anti-symmetric part, respectively.
Taking into account  Eq.(\ref{eq:Bdot1}), Eq.(\ref{eq:th-dot}) and 
Eq.(\ref{eq:h-dot}), we finally obtain 
%%%%%%%%%%%%%%%%%%%%%%%%%%%%%%%%%%%%
%%%%%%%%%%%%%%%%%%%%%%%%%%%%%%%%%%%%
\Bea
&& \stackrel{\circ}{\s}{}_{ab}
=  
    \half \gddot \s_{ab} + \frac{1}{n-1}\, \ul{\g} \cdot \s\,  h_{ab}
+  \widetilde{(\ul{\g} \s)} {}_{\ol{a}\ol{b}} 
+ \( \ul{\g} \om \){}_{\ol{a}\ol{b}}   
\nonumber \\
&& \qquad
+ \half \ddt \widetilde{\ul{\g}}{}_{ab}  
+ \widetilde{ \bmD_{\ol{a}} \nu_{\ol{b}}} 
+ 2 \xi_{\ol{a}} (\s \nu){}_{\ol{b}} 
\nonumber \\
&& \qquad 
-\xi_{\ol{a}}\(  
   \frac{\th}{n-1}  \ul{\g}\, {}_{\ol{b}} + (\s \ul{\g})_{\ol{b}}
-(\om \ul{\g})_{\ol{b}} \)
    \ \ ,
\label{eq:sigma-dot}  \\
%%%%%%%%%%%%%%%%%%%%%%%%%%%%%%%%%%%%
%\Bea
&& \stackrel{\circ}{\om}{}_{ab}
= - \half \gddot \om_{ab} 
-2 \xi_{\ul{a}} \( \om \nu \)_{\ul{b}}- \bmD_{\ul{a}} \nu_{\ul{b}} 
\nonumber \\ 
&& \qquad
+ \xi_{\ul{a}} 
\( \frac{\th}{n-1} \ul{\g}\, {}_{\ul{b}} 
+ (\s \ul{\g})_{\ul{b}} - (\om \ul{\g})_{\ul{b}}  \)  \ \ ,
\label{eq:omega-dot}
\Eea
%%%%%%%%%%%%%%%%%%%%%%%%%%%%%%%%%%%%%%
%%%%%%%%%%%%%%%%%%%%%%%%%%%%%%%%%%%%%%
where the second term on the R.H.S. of Eq.(\ref{eq:sigma-dot}) is the 
only trace-part of $\stackrel{\circ}{\sigma}{}_{ab}$, which vanishes 
for the case of the conformal geometry variation 
(${\ul{\g}}{}_{ab} \propto h_{ab} $). 
The last lines of Eq.(\ref{eq:sigma-dot}) and Eq.(\ref{eq:omega-dot})
also vanish for the case of the conformal geometry variation 
($\ul{\g}\, {}^a =0$). (The case of conformal geometry variations shall be analyzed in detail in Sec.\ref{sec:Conformal}.)  

Let us note  that $\om_{ab}=0$ {\sl iff} 
the timelike geodesic congruence $\C$ in question allows smooth $(n-1)$-dimensional spatial sections (see {\it Appendix} {\ref{app:formulas}} for more details).
Once $\om_{ab}=0$ is satisfied, it is satisfied all the way along the geodesic $\g$ as 
is seen by the third equation in Eq.(\ref{eq:Raych}). 
When discussing gravitational collapses, thus,  it is mostly 
 assumed $\om_{ab}=0$, since we are usually  interested in gravitational collapses 
started from ordinary, mild  initial conditions satisfying $\om_{ab}=0$.

When considering the geometry variations, it is thus a reasonable assumption 
that the class of variations we consider  retains the property $\om_{ab}=0$.
As discussed above, 
it corresponds to restricting the variations within collapsing geometries  
started from ordinary, mild  initial conditions.
From Eq.(\ref{eq:omega-dot}),  this condition for the geometry variation 
$\g_{ab}$ should be 
$\stackrel{\circ}{\om}{}_{ab}=0$,  yielding 
%%%%%%%%%%%%%%%%%%%%%%%%%%%%%%%%%%%%%%
%%%%%%%%%%%%%%%%%%%%%%%%%%%%%%%%%%%%%%
\Beq
\bmD_{\ul{a}} \nu_{\ul{b}}
- \xi_{\ul{a}} 
\( \frac{\th}{n-1} \ul{\g}\, {}_{\ul{b}} 
+ (\s \ul{\g})_{\ul{b}}  \) =0 \ \ .
\label{eq:omega-condition}
\Eeq
%%%%%%%%%%%%%%%%%%%%%%%%%%%%%%%%%%%
%%%%%%%%%%%%%%%%%%%%%%%%%%%%%%%%%%%
With the help of Eq.(\ref{eq:nusolve}), then, Eq.(\ref{eq:omega-condition}) is 
understood as the condition restricting  the geometry variation $\g_{ab}$    
 to those that preserve the well-behaved nature of the geodesic congruence 
described by $\om_{ab}=0$.  

\subsection{\label{subsec:dBdt-dot} 
Deviation of $\ddt B_{ab}$ due to geometry variations}

Noting that 
%%%%%%%%%%%%%%%%%%%%%%%%%%%%%%%%%%%%
%%%%%%%%%%%%%%%%%%%%%%%%%%%%%%%%%%%%
\[
{R_{acb}}^d(\l) ={R_{acb}}^d 
-2 \nb_{\ul{a}} {{\C}_{{\ul{c}}b}}^d (\l) 
+2  {{\C}_{\ul{a}b}}^e (\l) {{\C}_{\ul{c}e}}^d (\l) \ \ , 
\]
%%%%%%%%%%%%%%%%%%%%%%%%%%%%%%%%%%%%
%%%%%%%%%%%%%%%%%%%%%%%%%%%%%%%%%%%%
we get
%%%%%%%%%%%%%%%%%%%%%%%%%%%%%%%%%%%%
%%%%%%%%%%%%%%%%%%%%%%%%%%%%%%%%%%%%
\Beq
 {{\stackrel{\circ}{R}}_{acb}}{}^d
=-2 \nb_{\ul{a}} {{\stackrel{\circ}{\C}{}^d}_{\ul{c}b}} 
=-2 \nb_{\ul{a}} {\G^d}_{\ul{c}b}(\g)\ \ ,
\label{eq:Rcdot}
\Eeq
%%%%%%%%%%%%%%%%%%%%%%%%%%%%%%%%%%%%
%%%%%%%%%%%%%%%%%%%%%%%%%%%%%%%%%%%%
where
%%%%%%%%%%%%%%%%%%%%%%%%%%%%%%%%%%%%
%%%%%%%%%%%%%%%%%%%%%%%%%%%%%%%%%%%%
\Beaa
&& \G_{cab} (\g) \equiv \frac{1}{2}
                      (\nb_a \g_{bc} + \nb_b \g_{ac}  -\nb_c \g_{ab})\ \ , \\
&& {\G^c}_{ab} (\g)\equiv \frac{1}{2}g^{cd} \G_{dab} (\g) \ \ .
\Eeaa
%%%%%%%%%%%%%%%%%%%%%%%%%%%%%%%%%%%%
%%%%%%%%%%%%%%%%%%%%%%%%%%%%%%%%%%%%
Now taking the $\circ$-derivative on both sides of
 Eq.(\ref{eq:dBdt}) with $g_{ab}$ replaced by  $g_{ab}^{(\l)}$,  we get   
%%%%%%%%%%%%%%%%%%%%%%%%%%%%%%%%%%%%
%%%%%%%%%%%%%%%%%%%%%%%%%%%%%%%%%%%%
\Bea
&& \(\ddt B_{ab}\)^\circ 
=  - {\stackrel{\circ}{B}}_{ac}{B^c}_b - B_{ac}{{\stackrel{\circ}B}{}^c}_{b} 
-{{\stackrel{\circ}{R}}_{acb}}{}^d \xi^c \xi_d \nonumber \\
&& \qquad \qquad \qquad 
 -{R_{acb}}^d {\stackrel{\circ}{\xi}}{}^c \xi_d 
-{R_{acb}}^d \xi^c   {\stackrel{\circ}{\xi}}{}_d 
\ \ .
\label{eq:DelBdot}
\Eea
%%%%%%%%%%%%%%%%%%%%%%%%%%%%%%%%%%%%
%%%%%%%%%%%%%%%%%%%%%%%%%%%%%%%%%%%%
With the help of 
Eq.(\ref{eq:xidot}),  Eq.(\ref{eq:Bdot1}), Eq.(\ref{eq:Bdot2}) 
 and Eq.(\ref{eq:Rcdot}), we can 
rewrite  Eq.(\ref{eq:DelBdot}), yielding 
%%%%%%%%%%%%%%%%%%%%%%%%%%%%%%%%%%%%
%%%%%%%%%%%%%%%%%%%%%%%%%%%%%%%%%%%%
\Bea
&& \( \ddt B_{ab} \)^\circ 
= 2 \nb_{\ul{a}} \G_{{}^\Dot b \ul{c} }  \xi^c 
      -\frac{1}{2} \left\{ \frac{d \ul{\g}}{d\t}, B   \right\}_{ab} \nonumber \\
&& \ \    -B_{ac} (\bmD_b \nu^c +  \xi_b (B\nu)^c)
   -B_{cb} (\bmD^c \nu_a +  \xi_a (\nu B)^c)   \nonumber \\
&& \ \    -R_{acbd}(2 \nu^{\ol{c}} \xi^{\ol{d}} + \xi^c \ul{\g}^d ) 
   \nonumber \\
&&  -\( (\g B)_{\ol{a}\ol{c}} +\frac{1}{2}\gddot B_{ca}  \) {B^c}_b 
  -{B_a}^c \( (\g  B)_{\ul{b}\ul{c}} +\frac{1}{2}\gddot B_{bc}  \) \nonumber \\
&& + \left\{  B, \bmD \ul{\ast}  \ul{\g}\ul{\ast} 
           + \xi \ul{\ast} (\ul{\g} B)\ul{\ast} \right\}_{ab} \ \ .
\label{eq:DelBdot2}
\Eea
%%%%%%%%%%%%%%%%%%%%%%%%%%%%%%%%%%%%
%%%%%%%%%%%%%%%%%%%%%%%%%%%%%%%%%%%%
Here  $\G_{{}^\Dot bc} \equiv \G_{dbc} \xi^d $ (see Sec.\ref{subsec:notations} 
{\bf (iv)}); 
$\left\{A , B \right\}_{ab}\equiv \left(AB + BA \right)_{ab}$, the anti-commutator between $A$ and $B$, so that  
$\left\{  B, \bmD \ul{\ast}  \ul{\g}\ul{\ast} 
           + \xi \ul{\ast} (\ul{\g} B)\ul{\ast} \right\}_{ab}$ 
is the anti-commutator between $B_{ab}$ and 
 $ \bmD_{\ul{a}}  \ul{\g}_{\ul{b}} 
           + \xi_{\ul{a}} (\ul{\g} B)_{\ul{b}} $.

With some computations, it is straightforward to get
%%%%%%%%%%%%%%%%%%%%%%%%%%%%%%%%%%%%
%%%%%%%%%%%%%%%%%%%%%%%%%%%%%%%%%%%%
\Bea
&& 2 \nb_{\ul{a}} \G_{{}^\Dot b\ul{c}} (\g)  \xi^c  \nonumber \\
&& = \half \dddtt \ul{\g}{}_{ab} - \half \( \ddt \gddot\)  B_{ba} 
    +\half \bmD_a \bmD_b \gddot \nonumber \\
 && + (\ul{\g}{}_{cd} + \gddot h_{cd}) {B^d}_b {B^c}_a 
 +\ddt \left( 
( \ul{\g}{}_{\ol{a} c} + \gddot h_{\ol{a}c} ) {B^c}_{\ol{b}}\right) 
   \nonumber \\
&& -2 \bmD_{\ol{a}} \ul{\g}{}_c\, {B^c}_{\ol{b}} 
     -\ul{\g}{}_c\, \bmD_a {B^c}_b 
     -  \ul{\g}{}_c \left(\ddt {B^c}_a \right)  \xi_b  \nonumber \\
&& 
 - ( \ul{\g} BB )_a \xi_b - \ddt \bmD_{\ol{a}}  \ul{\g}{}_{\ol{b}} \ \ .
\label{eq:DelGamma}
\Eea
%%%%%%%%%%%%%%%%%%%%%%%%%%%%%%%%%%%%
%%%%%%%%%%%%%%%%%%%%%%%%%%%%%%%%%%%%
 
It is worth noting that, in case of  the conformal geometry variation 
described by $\g_{ab} = 2f g_{ab}$ ($f$ is any smooth function), 
it follows that $\ul{\g}{}_{ab} + \gddot h_{ab}=0$ and $\ul{\g}_a =0$ so that 
all the lines  but the first one on the R.H.S. of Eq.(\ref{eq:DelGamma}) 
vanish and the expression gets greatly simplified. 

Combing Eq.(\ref{eq:DelBdot2})  with  Eq.(\ref{eq:DelGamma}), we finally obtain
%%%%%%%%%%%%%%%%%%%%%%%%%%%%%%%%%%%%
%%%%%%%%%%%%%%%%%%%%%%%%%%%%%%%%%%%%
\Bea
&& \( \ddt B_{ab} \)^\circ
    \nonumber \\
&& = \half \dddtt {\ul{\g}} {}_{ab}
   -\half \left\{ \frac{d \ul{\g}}{d\t}, B   \right\}_{ab}
  -\half \( \ddt \gddot\) B_{ba} +  \half \bmD_a \bmD_b \gddot 
    \nonumber \\
&& -\left( B_{ac} \bmD_b \nu^c    +  \bmD_c \nu_a {B^c}_b \right)
    - \left(  \xi_a (\nu B B)_b  + (BB\nu)_a \xi_b \right) 
    \nonumber \\
&& -R_{acbd}(2 \nu^{\ol{c}} \xi^{\ol{d}} + \xi^c \ul{\g}^d ) 
    \nonumber \\
&& 
+ (\ul{\g}{}_{cd} + \gddot h_{cd}) {B^d}_b {B^c}_a
 +\ddt \left( 
( \ul{\g}{}_{\ol{a} c} + \gddot h_{\ol{a}c} ) {B^c}_{\ol{b}}\right) 
    \nonumber \\
&& -2 \bmD_{\ol{a}} \ul{\g}{}_c\, {B^c}_{\ol{b}} 
     -\ul{\g}{}_c\, \bmD_a {B^c}_b 
     -  \ul{\g}{}_c \left(\ddt {B^c}_a \right)  \xi_b 
    \nonumber \\
&& 
- ( \ul{\g} BB )_a \xi_b - \ddt \bmD_{\ol{a}}  \ul{\g}{}_{\ol{b}}
    \nonumber \\
&&
-\( (\g B)_{\ol{a}\ol{c}} +\frac{1}{2}\gddot B_{ca}  \) {B^c}_b 
  -{B_a}^c \( (\g  B)_{\ul{b}\ul{c}} +\frac{1}{2}\gddot B_{bc}  \) 
\nonumber \\
&& + \left\{  B, \bmD \ul{\ast}  \ul{\g}\ul{\ast}  \right\}_{ab}
       +\half \( B (\ul{\g}B) \){}_a\, \xi_b - \half \xi_a\> \( \ul{\g} BB \){}_b\ \ .
\label{eq:DelBdotfinal}
\Eea
%%%%%%%%%%%%%%%%%%%%%%%%%%%%%%%%%%%%
%%%%%%%%%%%%%%%%%%%%%%%%%%%%%%%%%%%%
Here we  note that   $( \ul{\g} BB )_a = \ul{\g}{}_b {B^b}_c {B^c}_a $ and 
$\( B (\ul{\g}B) \){}_a=B_{ab} \ul{\g}{}_c  B^{cb}$, faithfully following the 
notation rules introduced in Sec.\ref{subsec:notations}.  

We note that 
$\( \ddt B_{ab} \)^\circ$  as well as  ${\stackrel{\circ}{B}}_{ab}$ 
is not a spatial tensor.

In the case of the conformal geometry variation, $\g_{ab}=2 f g_{ab}$, only the 
first three lines on the R.H.S. of Eq.(\ref{eq:DelBdotfinal}) remain. 

To get explicit expressions for 
$\( \ddt \th \)^\circ$, $\(\ddt \s_{ab} \)^\circ$ and $\( \ddt \om_{ab} \)^\circ$ 
from  Eq.(\ref{eq:DelBdotfinal}), one can follow the similar procedures 
to get Eq.(\ref{eq:th-dot}), Eq.(\ref{eq:sigma-dot}) and Eq.(\ref{eq:omega-dot}) 
from Eq.(\ref{eq:Bdot1}). 
However, Eq.(\ref{eq:DelBdotfinal}) might be enough for the moment and 
we shall give the explicit expressions for the case of the 
conformal geometry variations below. 

%%%%%%%%%%%%%%%%%%%%%%%%%%%%%%%%%%%
%%%%%%%%%%%%%%%%%%%%%%%%%%%%%%%%%%%

\section{\label{sec:Conformal}Conformal  variations to the geometry}
\subsection{\label{subsec:ConformalFormulas}Basic formulas for the conformal variations to the geometry}

From now on, we confine ourselves to the cases of 
the conformal variations to the geometry  described as 
%%%%%%%%%%%%%%%%%%%%%%%%%%%%%%%%%
%%%%%%%%%%%%%%%%%%%%%%%%%%%%%%%%%%
\Beq
g_{ab}^{(\l)}(x)=\mathrm{e}^{2 \l  f(x)} g_{ab}(x)\ \ ,
\label{eq:conformal}
\Eeq
%%%%%%%%%%%%%%%%%%%%%%%%%%%%%%%%%
%%%%%%%%%%%%%%%%%%%%%%%%%%%%%%%%%%
 where $f(x)$ is an arbitrary smooth function. 

According to the first equation in Eq.(\ref{eq:g-dot}), we then get 
%%%%%%%%%%%%%%%%%%%%%%%%%%%%%%%%%
%%%%%%%%%%%%%%%%%%%%%%%%%%%%%%%%%%
\Beq
\g_{ab}(x)=2f(x) g_{ab}(x) \ \ .
\label{eq:gC}
\Eeq
%%%%%%%%%%%%%%%%%%%%%%%%%%%%%%%%%
%%%%%%%%%%%%%%%%%%%%%%%%%%%%%%%%%%
Based on Eq.(\ref{eq:gC}),   we further get
%%%%%%%%%%%%%%%%%%%%%%%%%%%%%%%%%
%%%%%%%%%%%%%%%%%%%%%%%%%%%%%%%%%%
\Bea
&&  \ul{\g}{}_{ab}= 2f h_{ab}\ \ , 
\ \ \ul{\g}\equiv \ul{\g}{{}_a}^a=2(n-1) f \ \ , 
\ \ \widetilde{\ul{\g}}{}_{ab}=0 \nonumber \\
&& \g_a = 2f \xi_a \ \ , \ \ \gddot = -2f \ \ , \ \  \ul{\g}{}_a = 0  \ \ , 
 \nonumber \\
&&  \ul{\g}{}_{ab} + \gddot h_{ab}=0 \ \ .
\label{eq:g-relC} 
\Eea
%%%%%%%%%%%%%%%%%%%%%%%%%%%%%%%%%
%%%%%%%%%%%%%%%%%%%%%%%%%%%%%%%%%%

Then, the key equation Eq.(\ref{eq:keyeq})  reduces to 
%%%%%%%%%%%%%%%%%%%%%%%%%%%%%%%%%%%
%%%%%%%%%%%%%%%%%%%%%%%%%%%%%%%%%%
\Beq
\frac{d\nu^a}{d\t} + {B^a}_b \nu^b = - \bmD^a f \ \ \  ,
\label{eq:keyeqC}
\Eeq
%%%%%%%%%%%%%%%%%%%%%%%%%%%%%%%%%%%%%%
%%%%%%%%%%%%%%%%%%%%%%%%%%%%%%%%%%%%%
or, in the form of Eq.(\ref{eq:Lop}), 
%%%%%%%%%%%%%%%%%%%%%%%%%%%%%%%%%%%%%%
%%%%%%%%%%%%%%%%%%%%%%%%%%%%%%%%%%%%%%
\Beq
{\mathcal{L}^a}_b \nu^b = - \bmD^a f \ \ , 
\label{eq:LopC}
\Eeq
%%%%%%%%%%%%%%%%%%%%%%%%%%%%%%%%%%%%%%
%%%%%%%%%%%%%%%%%%%%%%%%%%%%%%%%%%%%%%
where ${\mathcal{L}^a}_b$ is  given by  Eq.(\ref{eq:Leq}), which is invertible 
as discussed just after  Eq.(\ref{eq:Leq}). Thus 
Eq.(\ref{eq:keyeqC}) or Eq.(\ref{eq:LopC}) can be formally solved 
for $\nu^a$ as 
%%%%%%%%%%%%%%%%%%%%%%%%%%%%%%%%%%%%
%%%%%%%%%%%%%%%%%%%%%%%%%%%%%%%%%%%%%
\Beq
  \nu^a = - {\(\mathcal{L}^{-1}\)^a}_b \bmD^b f
= - \Biggl[
\(\ddt\bm{1} +  \bm{B} \)^{-1} \bmD f
\Biggr]^a \ \ .
\label{eq:nusolveC}
\Eeq
%%%%%%%%%%%%%%%%%%%%%%%%%%%%%%%%%%%%%
%%%%%%%%%%%%%%%%%%%%%%%%%%%%%%%%%%%%% 
Now, when $\g_{ab} = 2f g_{ab}$,  
Eq.(\ref{eq:Bdot1})   reduces to 
%%%%%%%%%%%%%%%%%%%%%%%%%%%%%%%%%%%%
%%%%%%%%%%%%%%%%%%%%%%%%%%%%%%%%%%%%
\Beq
 {\stackrel{\circ}{B}}_{ab} 
                   = f B_{ab} + \bmD_b \nu_a + \frac{df}{d\t} h_{ab}  
                      +(B \nu)_a \xi_b + \xi_a (\nu B)_b\ \ .
\label{eq:BdotC}
\Eeq
%%%%%%%%%%%%%%%%%%%%%%%%%%%%%%%%%%%%
%%%%%%%%%%%%%%%%%%%%%%%%%%%%%%%%%%%%
From Eq.(\ref{eq:BdotC}) one can extract equations for $\stackrel{\circ}{\th}$, 
$\stackrel{\circ}{\s}{}_{ab}$ and $\stackrel{\circ}{\om}{}_{ab}$ as before, 
which of course coincide with the equations reduced directly from 
Eq.(\ref{eq:th-dot}), Eq.(\ref{eq:sigma-dot}) and Eq.(\ref{eq:omega-dot}) 
for the case $\g_{ab} = 2f g_{ab}$.  
In order to reduce Eq.(\ref{eq:sigma-dot}) to the corresponding 
equation (Eq.(\ref{eq:th-s-om-dotC}) below), 
for instance, we note Eq.(\ref{eq:g-relC}) along with 
 $\ul{\g}\cdot \s = 2f {\s_a}^a =0$,  
$\widetilde{(\ul{\g} \s)} {}_{\ol{a}\ol{b}}= 2f \s_{ab}$,  
$\( \ul{\g} \om \){}_{\ol{a}\ol{b}}=2f \om_{\ol{a}\ol{b}}=0$ and so on.       
 We thus get 
%%%%%%%%%%%%%%%%%%%%%%%%%%%%%%%%%%%%
%%%%%%%%%%%%%%%%%%%%%%%%%%%%%%%%%%%%
\Bea
&& \stackrel{\circ}{\th} = -f  \th + (n-1) \frac{df}{d\t} + \bmD \cdot \nu \ \ ,
\nonumber \\
%%%%%%%%%%%%%%%%%%%%%%%%%%%%%%%%%%%%%%
%%%%%%%%%%%%%%%%%%%%%%%%%%%%%%%%%%%%
&& \stackrel{\circ}{\s}{}_{ab}
=  f  \s_{ab} + \widetilde{ \bmD_{\ol{a}} \nu_{\ol{b}}}
+  2 \xi_{\ol{a}} (\s \nu){}_{\ol{b}} \ \ ,
\nonumber  \\
%%%%%%%%%%%%%%%%%%%%%%%%%%%%%%%%%%%%
&& \stackrel{\circ}{\om}{}_{ab}
= f \om_{ab} - \bmD_{\ul{a}} \nu_{\ul{b}}
-2 \xi_{\ul{a}} \( \om \nu \)_{\ul{b}}
  \ \ . 
\label{eq:th-s-om-dotC}
\Eea
%%%%%%%%%%%%%%%%%%%%%%%%%%%%%%%%%%%%%%
%%%%%%%%%%%%%%%%%%%%%%%%%%%%%%%%%%%%%%
As for  $\stackrel{\circ}{\s}{}_{ab}$, 
comparing  Eq.(\ref{eq:sigma-dot}) and Eq.(\ref{eq:th-s-om-dotC}), 
we see that the only trace-part  on the R.H.S. of Eq.(\ref{eq:sigma-dot}), 
$\frac{1}{n-1} \ul{\g}\cdot \s h_{ab} $, vanishes in the conformal cases, 
so that the latter equation becomes completely trace-free. We also note that 
the last lines in Eq.(\ref{eq:sigma-dot}) and Eq.(\ref{eq:omega-dot})  
vanish in the conformal cases, due to $\ul{\g}{}_a =0$. 

Furthermore the condition for preserving $\om_{ab}=0$ 
(Eq.(\ref{eq:omega-condition})) reduces to a simple form
%%%%%%%%%%%%%%%%%%%%%%%%%%%%%%%%%%%
%%%%%%%%%%%%%%%%%%%%%%%%%%%%%%%%%%%
\Beq
\bmD_{\ul{a}} \nu_{\ul{b}}=0\ \ ,
\label{eq:omega-condition2}
\Eeq
%%%%%%%%%%%%%%%%%%%%%%%%%%%%%%%%%%%
%%%%%%%%%%%%%%%%%%%%%%%%%%%%%%%%%%%
which means that the geodesic shift $\nu^a$  should not form any rotation 
in the $(n-1)$-dimensional section orthogonal to the flow-line. 
A typical situation is that $\nu^a$ is described as a gradient of some scalar function 
as is the case analyzed in the next subsection  Sec.\ref{subsec:FRW} 
(see Eq.(\ref{eq:nusolveFRW}) below).

As for the equation for $\( \ddt B_{ab} \)^\circ$, Eq.(\ref{eq:DelBdotfinal})
also simplifies drastically for the conformal cases. Indeed, 
only the first three lines on the R.H.S. of  Eq.(\ref{eq:DelBdotfinal}) remain, yielding
%%%%%%%%%%%%%%%%%%%%%%%%%%%%%%%%%%%%
%%%%%%%%%%%%%%%%%%%%%%%%%%%%%%%%%%%%
\Bea
&& \( \ddt B_{ab} \)^\circ
    \nonumber \\
&& =  \frac{d^2 f}{d\t^2} h_{ab} - \bmD_a \bmD_b f 
           -2 \frac{df}{d\t} B_{ab} + \frac{df}{d\t} B_{ba} 
    \nonumber \\
&& -\Bigl( B_{ac} \bmD_b \nu^c    +  \bmD_c \nu_a {B^c}_b \Bigr)
    - \Bigl(  \xi_a (\nu B B)_b  + (BB\nu)_a \xi_b \Bigr) 
    \nonumber \\
&& -2R_{acbd} \nu^{\ol{c}} \xi^{\ol{d}} \ \ .
\label{eq:DelBdotfinalC}
\Eea
%%%%%%%%%%%%%%%%%%%%%%%%%%%%%%%%%%%%
%%%%%%%%%%%%%%%%%%%%%%%%%%%%%%%%%%%%
From Eq.(\ref{eq:DelBdotfinalC}),  
we can extract  the formulas for 
$\(\ddt \th \)^\circ$, $\(\ddt \s_{ab} \)^\circ$ and
 $\(\ddt \om_{ab} \)^\circ$.

First of all, we note that 
%%%%%%%%%%%%%%%%%%%%%%%%%%%%%%%%%%
%%%%%%%%%%%%%%%%%%%%%%%%%%%%%%%%%%
\Beaa
&& \(\frac{d \th}{d\t}\)^\circ 
= \frac{d}{d\l}_{|\l=0}\Biggl(g^{ab(\l)} \ddt B_{ab}(\l)\Biggr) \\
&& \qquad = \stackrel{\circ}{g}{}^{ab} \ddt B_{ab} +g^{ab} \( \ddt B_{ab} \)^\circ\ \ .
\Eeaa
%%%%%%%%%%%%%%%%%%%%%%%%%%%%%%%%%%
%%%%%%%%%%%%%%%%%%%%%%%%%%%%%%%%%%
Noting  that $\stackrel{\circ}{g}{}^{ab}= -2f g^{ab}$ due to 
Eq.(\ref{eq:g-dot2})  and Eq.(\ref{eq:gC}), 
the first term in the last line reduces to $-2f \frac{d\th}{d\t}$.
With the help of Eq.(\ref{eq:Raych}) and  Eq.(\ref{eq:DelBdotfinalC}), then,  we get
%%%%%%%%%%%%%%%%%%%%%%%%%%%%%%%%%%
%%%%%%%%%%%%%%%%%%%%%%%%%%%%%%%%%%
\Bea
&& \(\frac{d \th}{d\t}\)^\circ = 
 \frac{2f}{n-1}\ \th^2  
- \( \frac{df}{d\t}    + \frac{2}{n-1} \bmD \cdot \nu \) \th 
\nonumber \\
&& \qquad
+(n-1) \frac{d^2 f}{d\t^2} - \bmD \cdot \bmD f  
+ 2f R_{ {}^\Dot {}^\Dot }  
-2R_{{}^\Dot } \cdot \nu 
\nonumber \\
&& \qquad 
+ 2f \s \cdot \s - 2f \om \cdot \om 
- 2\s \cdot \bmD  \nu - 2\om \cdot \bmD  \nu
 \ \ . 
\label{eq:Dth-dotC}
\Eea
%%%%%%%%%%%%%%%%%%%%%%%%%%%%%%%%%%
%%%%%%%%%%%%%%%%%%%%%%%%%%%%%%%%%%

By the similar argument as in Eq.(\ref{eq:s-om-symmetry}), we see that 
$\(\ddt \s_{ab} \)^\circ$ and $\(\ddt \om_{ab} \)^\circ$ 
are symmetric and anti-symmetric in the indices, respectively,  due to 
the linearity of the $\circ$-derivative and $\ddt$. 
Thus $\(\ddt \om_{ab} \)^\circ$ is manifestly trace-free. 
Furthermore $\(\ddt \s_{ab} \)^\circ$ is also trace-free 
in the conformal cases, which is shown as
%%%%%%%%%%%%%%%%%%%%%%%%%%%%%%%%%%%%
%%%%%%%%%%%%%%%%%%%%%%%%%%%%%%%%%%%%
\Beaa
&& g^{ab} \(\ddt \s_{ab} \)^\circ 
=  \( g^{ab} \ddt \s_{ab}  \)^\circ   
-  \stackrel{\circ}{g}{}^{ab} \ddt \s_{ab} \\
&& \qquad \qquad \qquad = 2f g^{ab} \ddt \s_{ab} =0\ \ ,
\Eeaa
%%%%%%%%%%%%%%%%%%%%%%%%%%%%%%%%%%%%
%%%%%%%%%%%%%%%%%%%%%%%%%%%%%%%%%%%%
where we have used the fact that $\ddt \s_{ab}$ is trace-free  
(as is seen by Eq.(\ref{eq:Raych})).

Thus, by extracting the trace-free part of Eq.(\ref{eq:DelBdotfinalC}) and 
dividing it into the symmetric part and the anti-symmetric part,  we get
%%%%%%%%%%%%%%%%%%%%%%%%%%%%%%%%%%%%
%%%%%%%%%%%%%%%%%%%%%%%%%%%%%%%%%%%%
\Bea
&& \(\ddt \s_{ab} \)^\circ
= -\frac{df}{d\t} \s_{ab} - \widetilde{\bmD_{\ol{a}} \bmD_{\ol{b}}} f 
\nonumber \\
&& -\frac{2 \th}{n-1}  \widetilde{ \bmD_{\ol{a}} \nu_{\ol{b}} }
-\widetilde{ \left\{ \s , \bmD \nu  \right\} }_{\ol{a} \ol{b}}
+\widetilde{ \left\{ \om , \bmD \nu  \right\} }_{\ol{a} \ol{b}}
\nonumber \\
&& -\frac{2 \th^2}{(n-1)^2} \xi_{\ol{a}} \nu_{\ol{b}}
-\frac{4 \th}{n-1} \xi_{\ol{a}} (\s \nu)_{\ol{b}}
-2 \xi_{\ol{a}} (\s \s \nu)_{\ol{b}} -2 \xi_{\ol{a}} (\om \om \nu)_{\ol{b}}
\nonumber \\
&&
 -2(R_{acbd} -\frac{1}{n} R_{cd}g_{ab}) \nu^{\ol{c}} \xi^{\ol{d}}\ \ ,
\label{eq:Ds-dotC} 
\Eea
%%%%%%%%%%%%%%%%%%%%%%%%%%%%%%%%%%%%
%%%%%%%%%%%%%%%%%%%%%%%%%%%%%%%%%%%%
and
%%%%%%%%%%%%%%%%%%%%%%%%%%%%%%%%%%%%
%%%%%%%%%%%%%%%%%%%%%%%%%%%%%%%%%%%%
\Bea
&& \(\ddt \om_{ab} \)^\circ
= -2 \frac{df}{d\t} \om_{ab}  
\nonumber \\
&&
 + \frac{2 \th}{n-1} \bmD_{\ul{a}} \nu_{\ul{b}} 
+ \left\{ \s , \bmD_{\ul{*}} \nu_{\ul{*}}  \right\}_{ab}
- \left[ \om , \bmD_{\ol{*}} \nu_{\ol{*}}  \right]_{ab}
\nonumber \\
&& 
+\frac{4 \th}{n-1} \xi_{\ul{a}} (\om \nu)_{\ul{b}}
+2 \xi_{\ul{a}} \left(\left\{ \s , \om \right\} \nu \right)_{\ul{b}} \ \ .
\label{eq:Dom-dotC} 
\Eea
%%%%%%%%%%%%%%%%%%%%%%%%%%%%%%%%%%%%
%%%%%%%%%%%%%%%%%%%%%%%%%%%%%%%%%%%%
We  see from  Eq.(\ref{eq:Dom-dotC}) and 
the third equation in Eq.(\ref{eq:th-s-om-dotC}) 
  that one can set $\om_{ab} \equiv 0$ provided 
 that the ``twist-less" condition Eq.(\ref{eq:omega-condition2}) is satisfied for 
the geometry variations.

We note that the arbitrary function $f(x)$ determines the geometry variation 
$\g_{ab}$ (Eq.(\ref{eq:gC})), which causes the geodesic shift 
(Eq.(\ref{eq:nusolveC})).
Then the two sets of equations (the equations in Eq.(\ref{eq:th-s-om-dotC}) and 
the equations 
Eq.(\ref{eq:Dth-dotC})-Eq.(\ref{eq:Dom-dotC}))   describe the linear responses of 
$\th$, $\s$ and $\om$ and those of  
$\frac{d\th}{d\t}$, $\ddt\s_{ab} $ and $\ddt\om_{ab}$  
  to the geometry variation caused by the arbitrary function $f$.

Since the function $f(x)$ is arbitrary, then, we can analyze in detail  the 
relation between the geometry variations and 
the changes in the focusing properties of 
the geodesic congruence 
 with the help of  these sets of equations.

\subsection{\label{subsec:FRW}Application to  the time-reversed Friedmann-Robertson-Walker model}

To get some concrete insights for the present framework, we here apply it to  
the case of the time-reversed Friedmann-Robertson-Walker (FRW) model. 
The FRW model describes  the expanding universe with an initial singularity, so that 
its  time-reversed version can be 
regarded as a simple spacetime model describing  a gravitational collapse with 
a final singularity. (See {\it Appendix} \ref{app:FRWapp} for the basic results of 
the FRW spacetime.)

The metric for the FRW spacetime is given in the co-moving coordinates as
%%%%%%%%%%%%%%%%%%%%%%%%%%%%%%%%%%%
%%%%%%%%%%%%%%%%%%%%%%%%%%%%%%%%%%%
\Bea
&& ds^2 
    = -dt^2 +a^2(t) \left( F^2(r) dr^2 + r^2 d\vartheta^2+ r^2\sin^2\vartheta
 d\varphi^2 \right)
\ \ , 
\nonumber \\
&& F(r) := \frac{1}{\sqrt{1-kr^2}} \qquad (k= -1,0,1)\ \ .
\label{eq:FRW}
\Eea
%%%%%%%%%%%%%%%%%%%%%%%%%%%%%%%%%%%
%%%%%%%%%%%%%%%%%%%%%%%%%%%%%%%%%%%
 Let us choose as a geodesic congruence $\C$ 
the set of standard geodesics compatible with  the  co-moving coordinates, 
described by $x^a(t) = {}^T (t\  r_0\  \vartheta_0\  \varphi_0 )$ with 
$r_0$, $\vartheta_0$ and $\varphi_0$ being some constants. Then in the 
present frame of coordinates it follows 
%%%%%%%%%%%%%%%%%%%%%%%%%%%%%%%%%%%
%%%%%%%%%%%%%%%%%%%%%%%%%%%%%%%%%%%
\Beaa
&& \xi^a={}^T (1\  0\ 0\ 0)
\ \ , \ \ 
\xi_a={}^T (-1\  0\ 0\ 0) \ \ , \\
&&
h_{ab}= {\rm diag} (0\ \  a^2F\ \   a^2 r^2\ \   a^2 r^2 \sin^2 \vartheta ) \ \ .
\Eeaa
%%%%%%%%%%%%%%%%%%%%%%%%%%%%%%%%%%%%
%%%%%%%%%%%%%%%%%%%%%%%%%%%%%%%%%%%%
It is then straightforward to get
%%%%%%%%%%%%%%%%%%%%%%%%%%%%%%%%%%%%
%%%%%%%%%%%%%%%%%%%%%%%%%%%%%%%%%%%%
\[
 B_{ab} = {\G^0}_{ab}= \frac{1}{2} g_{ab ,0} = \frac{\dot{a}}{a} h_{ab} \ \ , 
\]
%%%%%%%%%%%%%%%%%%%%%%%%%%%%%%%%%%%%
%%%%%%%%%%%%%%%%%%%%%%%%%%%%%%%%%%%%
which yields
%%%%%%%%%%%%%%%%%%%%%%%%%%%%%%%%%%%%
%%%%%%%%%%%%%%%%%%%%%%%%%%%%%%%%%%%%
\Beq
 \th = 3 \frac{\dot{a}}{a}\ \ , \ \ \s_{ab}=\om_{ab}=0\ \ .        
\label{eq:th-s-omFRW}
\Eeq
%%%%%%%%%%%%%%%%%%%%%%%%%%%%%%%%%%%
%%%%%%%%%%%%%%%%%%%%%%%%%%%%%%%%%%%
From Eq.(\ref{eq:Rabcdot}) and Eq.(\ref{eq:Rdotdot}) in 
{\it Appendix} \ref{app:FRWapp}, we see that  
%%%%%%%%%%%%%%%%%%%%%%%%%%%%%%%%%%
%%%%%%%%%%%%%%%%%%%%%%%%%%%%%%%%%%
\[
R_{a^\Dot  b^\Dot}   =-\frac{\ddot{a}}{a} h_{ab}\ \ , 
\ \  \widetilde{R}_{a^\Dot  b^\Dot} =0\ \ , \ \ 
R_{{}^\Dot  {}^\Dot} = -3\frac{\ddot{a}}{a}\ \ .
\]
%%%%%%%%%%%%%%%%%%%%%%%%%%%%%%%%%%%
%%%%%%%%%%%%%%%%%%%%%%%%%%%%%%%%%%%
As far as $\ddot{a} <0$ (a time-reversal invariant relation), thus, it follows 
that $R_{\Dot  \Dot} >0$. 

In the context of the {\it generic condition} 
for the singularity theorems~\cite{HawkingEllis, Wald}, which is stated as 
``each timelike geodesic contains at least one point at which 
$R_{a^\Dot  b^\Dot} \neq 0$",  
 the  trace-part of $R_{a^\Dot  b^\Dot}$ (i.e. $R_{\Dot  \Dot}$) 
is non-zero rather than  the trace-free part ($\widetilde{R}_{a^\Dot  b^\Dot}$). 
Because of this fact, $\frac{d\th}{d\t} < 0$ is guaranteed even though 
 $\s_{ab} \equiv 0$  (see Eq.(\ref{eq:Raych})). 

Then Eq.(\ref{eq:keyeqC}) in this case becomes
%%%%%%%%%%%%%%%%%%%%%%%%%%%%%%%%%%%
%%%%%%%%%%%%%%%%%%%%%%%%%%%%%%%%%%%
\Beq
\dot{\nu}^a + \frac{\dot{a}}{a}\nu^a= - \bmD^a f \ \ ,
\label{eq:keyeqFRW}
\Eeq
%%%%%%%%%%%%%%%%%%%%%%%%%%%%%%%%%%%
%%%%%%%%%%%%%%%%%%%%%%%%%%%%%%%%%%%
which can be explicitly solved  in the form of Eq.(\ref{eq:nusolveC}) as 
%%%%%%%%%%%%%%%%%%%%%%%%%%%%%%%%%%%
%%%%%%%%%%%%%%%%%%%%%%%%%%%%%%%%%%%
\Bea
&& \nu^a(t, \vec{x}) = \frac{a(t_0)}{a(t)}\nu^a(t_0, \vec{x}) 
- a(t) \bmD^a \psi(t, \vec{x}) \ \ , \ \ \nonumber \\
&& 
\psi(t, \vec{x}) \equiv \int _{t_0}^t \frac{f(t', \vec{x} )}{a(t')} dt'\ \ .
\label{eq:nusolveFRW}
\Eea
%%%%%%%%%%%%%%%%%%%%%%%%%%%%%%%%%%%
%%%%%%%%%%%%%%%%%%%%%%%%%%%%%%%%%%%
Here we should note that the operator $\bmD^a$ is $t$-{\it dependent}  
through $a(t)$, which is clearly seen  in the co-moving coordinates as 
 \[
\bmD^a = {}^T (\frac{1}{a^2(t)F^2(r)}\del_r \  \quad 
\frac{1}{a^2(t) r^2}\del_\vartheta\  \quad 
\frac{1}{a^2(t)r^2 \sin \vartheta}\del_\varphi   )\ \ .
\]
Thus Eq.(\ref{eq:nusolveFRW}) is expressed in a more explicit form
%%%%%%%%%%%%%%%%%%%%%%%%%%%%%
%%%%%%%%%%%%%%%%%%%%%%%%%%%%%
\Beq
\nu^a(t, \vec{x}) = \frac{a(t_0)}{a(t)}\nu^a(t_0, \vec{x}) 
- \frac{1}{a(t)} {\bmD_{[1]}}^a \psi(t, \vec{x})\ \ , 
\label{eq:nusolveFRWexplicit}
\Eeq
%%%%%%%%%%%%%%%%%%%%%%%%%%%%%
%%%%%%%%%%%%%%%%%%%%%%%%%%%%%
where ${\bmD_{[1]}}^a$ is $\bmD^a$ with $a(t) \equiv 1$ 
so that it is  $t$-independent.
(These two operators  are related by ${\bmD_{[1]}}^a \equiv a(t)^2 \bmD^a$. )   
One can then easily check that Eq.(\ref{eq:nusolveFRWexplicit}) is the solution 
for Eq.(\ref{eq:keyeqFRW}).

For simplicity, we shall set $\nu^a(t_0, \vec{x})=0$ below. 
Then Eq.(\ref{eq:th-s-om-dotC}) becomes 
%%%%%%%%%%%%%%%%%%%%%%%%%%%%%%%%%%%
%%%%%%%%%%%%%%%%%%%%%%%%%%%%%%%%%%%
\Bea
&& \stackrel{\circ}{\th} = - 3\frac{\dot{a}}{a}f   + 3 \frac{df}{d\t} 
- a(t) \bmD \cdot \bmD \psi(t, \vec{x}) \ \ ,  
\nonumber \\
%%%%%%%%%%%%%%%%%%%%%%%%%%%%%%%%%%%%%%
%%%%%%%%%%%%%%%%%%%%%%%%%%%%%%%%%%%%
&& \stackrel{\circ}{\s}{}_{ab}
= -a(t) \widetilde{ \bmD_{\ol{a}} \bmD_{\ol{b}}} \psi(t, \vec{x}) \ \ , 
\nonumber  \\
%%%%%%%%%%%%%%%%%%%%%%%%%%%%%%%%%%%%
&& \stackrel{\circ}{\om}{}_{ab} = 0 \ \ . 
\label{eq:dotCFRW}
\Eea
%%%%%%%%%%%%%%%%%%%%%%%%%%%%%%%%%%%
%%%%%%%%%%%%%%%%%%%%%%%%%%%%%%%%%%%
We note that Eq.(\ref{eq:dotCFRW}) describes   
how the focusing of geodesics starts to deviate from the one in the exact FRW 
spacetime:  
The conformal geometry variation induces the variations in   
 $\th$ and $\s_{ab}$.  In particular $\s_{ab}$ obtains a non-zero value 
 even though $\s_{ab}=0$ at the beginning.  
 Once $\s_{ab}$ gets non-zero, the first and the third terms on the R.H.S. of 
the second equation in Eq.(\ref{eq:th-s-om-dotC}) also cause 
the variation in $\s_{ab}$. On the other hand, $\om_{ab}=0$ 
is  preserved as far as the condition Eq.(\ref{eq:omega-condition2}) is satisfied.

Furthermore 
Eq.(\ref{eq:Dth-dotC}), Eq.(\ref{eq:Ds-dotC}) and Eq.(\ref{eq:Dom-dotC}) 
become
%%%%%%%%%%%%%%%%%%%%%%%%%%%%%%%%%%
%%%%%%%%%%%%%%%%%%%%%%%%%%%%%%%%%%
\Bea
&& \(\frac{d \th}{d\t}\)^\circ = 
3 \frac{d^2 f}{d\t^2}
- 3 \frac{\dot{a}}{a} \frac{df}{d\t}    
+ 6  \Bigl\{ \(\frac{\dot{a}}{a} \)^2   - \frac{\ddot{a}}{a} \Bigr\} f  
\nonumber \\
&& \qquad \qquad + 2 \dot{a} \bmD \cdot \bmD \psi   - \bmD \cdot \bmD f \ \ . 
\nonumber \\
%%%%%%%%%%%%%%%%%%%%%%%%%%%%%%%%%%%%
%%%%%%%%%%%%%%%%%%%%%%%%%%%%%%%%%%%%
&& \(\frac{d \s_{ab}}{d\t} \)^\circ
=  - \widetilde{\bmD_{\ol{a}} \bmD_{\ol{b}}} f 
 + 2 \dot{a}  \widetilde{ \bmD_{\ol{a}} \bmD_{\ol{b}} } \psi 
\nonumber \\
&& \qquad \qquad  + 2a  \Bigl\{ \(\frac{\dot{a}}{a} \)^2   - \frac{\ddot{a}}{a} 
\Bigr\}  \xi_{\ol{a}}\,  \bmD_{\ol{b}} \psi\ \ ,
\nonumber \\
%%%%%%%%%%%%%%%%%%%%%%%%%%%%%%%%%%%%
%%%%%%%%%%%%%%%%%%%%%%%%%%%%%%%%%%%%
&&  \(\frac{d \om_{ab}}{d\t} \)^\circ = 0\ \ . 
\label{eq:D-dotCFRW} 
\Eea
%%%%%%%%%%%%%%%%%%%%%%%%%%%%%%%%%%%%
%%%%%%%%%%%%%%%%%%%%%%%%%%%%%%%%%%%%

Here let us pay attention to the first equations in 
Eq.(\ref{eq:dotCFRW}) and  Eq.(\ref{eq:D-dotCFRW}) for 
$\dot{a}<0$, $\ddot{a}<0$ and $f<0$;
%%%%%%%%%%%%%%%%%%%%%%%%%%%%%%%%%%%%
%%%%%%%%%%%%%%%%%%%%%%%%%%%%%%%%%%%%

%%%%%%%%%%%%%%%%%%%%%%%%%%%%%%%%%%%%
%%%%%%%%%%%%%%%%%%%%%%%%%%%%%%%%%%%%
\Beaa
&& \stackrel{\circ}{\th} = - 3\frac{|\dot{a}|}{a}|f|   + 3 \frac{df}{d\t} 
- a(t) \int _0^t \frac{ \bmD \cdot \bmD f(t', \vec{x} )}{a(t')} dt' \ \ ,  
\nonumber \\
&& \(\frac{d \th}{d\t}\)^\circ = 
3 \frac{d^2 f}{d\t^2}
+ 3 \frac{|\dot{a}|}{a} \frac{df}{d\t}    
- 6  \Bigl\{ \(\frac{\dot{a}}{a} \)^2   + \frac{|\ddot{a}|}{a} \Bigr\} |f|  
\nonumber \\
&& \qquad \qquad 
- 2 |\dot{a}(t)|  \int _0^t \frac{ \bmD \cdot \bmD f(t', \vec{x} )}{a(t')} dt'
   - \bmD \cdot \bmD f \ \ .
%\label{eq:Dth-dotCFRW}
\Eeaa
%%%%%%%%%%%%%%%%%%%%%%%%%%%%%%%%%%%%
%%%%%%%%%%%%%%%%%%%%%%%%%%%%%%%%%%%%

The function $f(t,\vec{x})$ can be chosen arbitrarily.  
The above two equations  serve as  concrete examples for 
 the  general cases of the conformal 
variations, which is analyzed  in the next section. 

%%%%%%%%%%%%%%%%%%%%%%%%%%%%%%%%%%%%%%%
%%%%%%%%%%%%%%%%%%%%%%%%%%%%%%%%%%%%%%%
\section{\label{sec:Focusing}Influence of conformal geometry variations on 
the focusing properties  of  geodesic congruences}
%%%%%%%%%%%%%%%%%%%%%%%%%%%%%%%%%%%%%%%
%%%%%%%%%%%%%%%%%%%%%%%%%%%%%%%%%%%%%%%

We  now investigate the response of  the gravitational focusing properties  
of  geodesic congruences to conformal geometry variations. 
For this purpose, let us focus on   the behavior of the expansion $\th$. 

We first   recall how $\th$ behaves during the latest phase of gravitational 
contractions. 
We pay attention to the Raychaudhuri equation, 
the first equation in Eq.(\ref{eq:Raych}). 
As discussed at the end of Sec.\ref{subsec:expansion},  one can set 
${\om}_{ab}=0$ when the timelike geodesic congruence $\C$ is hypersurface 
orthogonal, 
which is usually assumed. Furthermore  
$R_{{}^\Dot {}^\Dot } >0$ is guaranteed   
assuming  that the {\it strong energy condition} is satisfied for the matter content. 
Thus 
%%%%%%%%%%%%%%%%%%%%%%%%%%%%%%%%%%%
%%%%%%%%%%%%%%%%%%%%%%%%%%%%%%%%%%%
\Beq
\frac{d \th}{d\t}  + \frac{1}{n-1}\ \th^2 \leq 0\ \ ,
\label{eq:th-ineq}
\Eeq  
%%%%%%%%%%%%%%%%%%%%%%%%%%%%%%%%%%%
%%%%%%%%%%%%%%%%%%%%%%%%%%%%%%%%%%%
which is equivalent to 
%%%%%%%%%%%%%%%%%%%%%%%%%%%%%%%%%%%
%%%%%%%%%%%%%%%%%%%%%%%%%%%%%%%%%%%
\[
\frac{d \th^{-1}}{d\t} \geq  \frac{1}{n-1} \ \ ,
\]
%%%%%%%%%%%%%%%%%%%%%%%%%%%%%%%%%%%
%%%%%%%%%%%%%%%%%%%%%%%%%%%%%%%%%%%
or
%%%%%%%%%%%%%%%%%%%%%%%%%%%%%%%%%%%
%%%%%%%%%%%%%%%%%%%%%%%%%%%%%%%%%%%
\[
\th (\t) \leq  \({\th (\t_0)}^{-1} + \frac{1}{n-1} (\t - \t_0)\)^{-1} \ \ .
\]  
%%%%%%%%%%%%%%%%%%%%%%%%%%%%%%%%%%%
%%%%%%%%%%%%%%%%%%%%%%%%%%%%%%%%%%%
Thus once $\th$ becomes negative at some $\t_0$ (i.e. $\th (\t_0) < 0$), then 
 $\th \rightarrow -\infty $ as the proper-time  tends to some finite value 
$\t$ satisfying $\t_0 < \t  \leq \t_0 + \frac{n-1}{|\th_0|}$~\cite{Wald}.

The function $f(x)$ is arbitrary as far as the consistency of  arguments are 
retained. In the present context of gravitational collapses, 
it is  reasonable to enforce  the strong energy condition 
 for the consistency with the argument just before Eq.(\ref{eq:th-ineq}). 
Indeed, taking into account the energy condition 
for restricting the class of functions to be considered 
is sometimes vital for 
describing the gravitational contractions correctly~\cite{Seriu}. 

Let us then pay attention to the strong energy condition, 
translated as $R_{{}^\Dot {}^\Dot} >0$,
and  derive a formula for $\(R_{{}^\Dot {}^\Dot}\)^\circ $. 
By the conformal transformation Eq.(\ref{eq:conformal}),  the Ricci tensor 
is transformed as~\cite{Wald}, 
%%%%%%%%%%%%%%%%%%%%%%%%%%%%%%%%%%%%
%%%%%%%%%%%%%%%%%%%%%%%%%%%%%%%%%%%%
\[
{R^{(\l)}}_{ab} = R_{ab}-\l (n-2)  \nb_a \nb_b f - \l g_{ab}  \D f + O(\l^2) \ \ .
%&& \qquad \qquad 
%+ \l^2 (n-2)  \nb_a f \nb_b f - \l^2 (n-2) g_{ab} \nb f \cdot \nb f \ \ .
\]
%%%%%%%%%%%%%%%%%%%%%%%%%%%%%%%%%%%%
%%%%%%%%%%%%%%%%%%%%%%%%%%%%%%%%%%%%
On the other hand, the tangent vector of a geodesic is transformed as
%%%%%%%%%%%%%%%%%%%%%%%%%%%%%%%%%%%%
%%%%%%%%%%%%%%%%%%%%%%%%%%%%%%%%%%%%
\[
\xi^{(\l)\ a} =(1-\l f)\xi^a + \l \nu^a + O(\l^2)\ \ ,
\]
%%%%%%%%%%%%%%%%%%%%%%%%%%%%%%%%%%%%
%%%%%%%%%%%%%%%%%%%%%%%%%%%%%%%%%%%%
where Eq.(\ref{eq:xidot}) and Eq.(\ref{eq:g-relC}) have been used. 
Thus we  obtain
%%%%%%%%%%%%%%%%%%%%%%%%%%%%%%%%%%%%
%%%%%%%%%%%%%%%%%%%%%%%%%%%%%%%%%%%%
\Beq
\(R_{{}^\Dot {}^\Dot}\)^\circ 
= -2f R_{{}^\Dot {}^\Dot} + \bmD \cdot \bmD f - \th \frac{df}{d\t} 
   -(n-1)\frac{d^2 f}{d\t^2} + 2 \nu \cdot R_{{}^\Dot}\ \ ,
\label{eq:Rdotdot-dot}
\Eeq
%%%%%%%%%%%%%%%%%%%%%%%%%%%%%%%%%%%%
%%%%%%%%%%%%%%%%%%%%%%%%%%%%%%%%%%%%
where Eq.(\ref{eq:Delta-Delta}) has been used and 
$\nu \cdot R_{{}^\Dot}  \equiv \nu^a R_{ab} \xi^b$ (following the notation rules 
in Sec.\ref{subsec:notations}). 

To retain the strong energy condition, it is reasonable to choose the function 
$f(x)$ which does not greatly depart from 
the condition $\(R_{{}^\Dot {}^\Dot}\)^\circ \geq 0$.

Now we pay attention to the first equation in Eq.(\ref{eq:th-s-om-dotC}),  and 
Eq.(\ref{eq:Dth-dotC}). 
We  consider the final phase of gravitational contractions where 
$\th$ is negative and $|\th| \D \t \gg 1$ according to the argument after 
Eq.(\ref{eq:th-ineq}).  Here $\D \t$ is the typical time-scale of  
changes of our concern.  
Let $\D \t_{\th}$ be the time-scale in which $\th$ changes substantially, then 
the above estimation implies $\D \t_{\th} \ll \D \t$. 
Next, the behavior of $f(x)$ is quite arbitrary 
 but at least $\D \t_{\th} \ll  \D T_f \ll  \D \t $ and 
$\D \t_{\th} \ll  \D X_f \ll  \D \t $ should satisfy in order for 
our arguments to be meaningful. 
In particular,  $\D \t$, the typical time-scale of our concern,  
should be large enough compared to the other scales 
to investigate the  effects of geometry {\it variations}.    
Thus we impose 
%%%%%%%%%%%%%%%%%%%%%%%%%%%%%%%%%%%%%
%%%%%%%%%%%%%%%%%%%%%%%%%%%%%%%%%%%%%
\Beq
\D \t_{\th} \ll   (\D T_f \ \ , \ \   \D X_f)  \ll \D \t \ \ ,
\label{eq:scales}
\Eeq
%%%%%%%%%%%%%%%%%%%%%%%%%%%%%%%%%%%%%
%%%%%%%%%%%%%%%%%%%%%%%%%%%%%%%%%%%%%
implying that 
\[
|\th| |\frac{df}{d\t}|   \gg    
  ( |\frac{d^2 f}{d\t^2}| \ \ , \ \  |\bmD \cdot \bmD f| ) \ \ .
\]
 
For simplicity, let us assume $f <0$ also, which roughly 
corresponds to the conformal geometry variations 
in favor of the geodesic focusing.

Now we  focus on  
the first equation in Eq.(\ref{eq:th-s-om-dotC}). 
Let us first analyze the importance of each term on the  R.H.S. of the equation.  
We see  from Eq.(\ref{eq:nusolveC}) (along with its counter-part for  the time-reversed FRW  model, Eq.(\ref{eq:nusolveFRW})) that 
%%%%%%%%%%%%%%%%%%%%%%%%%%%%%%%%%%%%%%
%%%%%%%%%%%%%%%%%%%%%%%%%%%%%%%%%%%%%%
\Beq
 \bmD \cdot \nu \sim   -\D \t  \bmD \cdot \bmD f   \sim - \frac{\D \t}{{\D X_f} ^2}f \ \ .
\label{eq:DnuEst}
\Eeq
%%%%%%%%%%%%%%%%%%%%%%%%%%%%%%%%%%%%%%
%%%%%%%%%%%%%%%%%%%%%%%%%%%%%%%%%%%%%%
Then we  estimate as 
%%%%%%%%%%%%%%%%%%%%%%%%%%%%%%%%%%%%%%%%
%%%%%%%%%%%%%%%%%%%%%%%%%%%%%%%%%%%%%%%%
\[
\frac{|\bmD \cdot \nu |}{ |f \th|} \sim \frac{\D \t \D \t_{\th}}{ {\D X_f}^2} 
\ \ \ , \ \ \ \frac{|\frac{df}{d\t} |}{ |f \th|} \sim \frac{\D \t_{\th}}{\D T_f}\ \ \ , \ \ \ \frac{|\frac{df}{d\t} |}{|\bmD \cdot \nu |} \sim 
\frac{{\D X_f}^2}{\D T_f \D \t }\ \ .
\]
%%%%%%%%%%%%%%%%%%%%%%%%%%%%%%%%%%%%%%%%
%%%%%%%%%%%%%%%%%%%%%%%%%%%%%%%%%%%%%%%%
There is no a priori  relation between $\D T_f$ and $\D X_f$. 
One reasonable situation is, however, that they are of similar magnitudes, e.g. 
$\D T_f \sim \D X_f$,  unless quite special situations are considered. 

Thus, along with Eq.(\ref{eq:scales}),  we see 
%%%%%%%%%%%%%%%%%%%%%%%%%%%%%%%%%%%%%%%%
%%%%%%%%%%%%%%%%%%%%%%%%%%%%%%%%%%%%%%%%
\Beq
(|f \th| \ , \  |\bmD \cdot \nu |) \ \ \gg \ \  |\frac{df}{d\t} |\ \ ,
\label{eq:estimation}
\Eeq
%%%%%%%%%%%%%%%%%%%%%%%%%%%%%%%%%%%%%%%%
%%%%%%%%%%%%%%%%%%%%%%%%%%%%%%%%%%%%%%%%
implying that the term $\bmD \cdot \nu$ is as important as 
the term $f \th$ and they are 
much more important than the term $\frac{df}{d\t}$.

Next let us search for the condition for $ \stackrel{\circ}{\th} <0 $ which implies 
the stronger focusing of geodesics due to the geometry variations. 
In later  phase of the geodesic congruence contractions, the threshold condition 
 $ \stackrel{\circ}{\th} =0 $ becomes
%%%%%%%%%%%%%%%%%%%%%%%%%%%%%%%%%%%
%%%%%%%%%%%%%%%%%%%%%%%%%%%%%%%%%%%
\Beq
  \frac{df}{d\t}+  \frac{|\th|}{n-1} f  =-  \frac{1}{n-1} \bmD \cdot \nu  \ \ ,
\label{eq:border1}
\Eeq
%%%%%%%%%%%%%%%%%%%%%%%%%%%%%%%%%%%
%%%%%%%%%%%%%%%%%%%%%%%%%%%%%%%%%%%
which along with $f(\t_0)=f_0$ can be solved as 
%%%%%%%%%%%%%%%%%%%%%%%%%%%%%%%%%%%
%%%%%%%%%%%%%%%%%%%%%%%%%%%%%%%%%%%
\Beaa
&& f(\t) = \(f_0 - \frac{1}{n-1} \int_{\t_0}^\t \bmD (\t'') \cdot \nu \ 
\mrE^{A(\t'' , \t_0)}\  d\t'' \) \ 
\mrE^{-A(\t, \t_0)}\ \ , \\
&& A(\t, \t_0) \equiv \frac{1}{n-1} \int_{\t_0}^\t |\th (\t')| d\t'\ \ .
\Eeaa
%%%%%%%%%%%%%%%%%%%%%%%%%%%%%%%%%%%
%%%%%%%%%%%%%%%%%%%%%%%%%%%%%%%%%%%
We note that 
\[
A(\t, \t_0) \sim  \frac{|\bar{\th}|}{n-1} \D \t  
\sim \frac{\D \t}{\D \t_{\th}}\ \ , 
\]
 where $\D \t =\t -\t_0$ and $\bar{\th}$ is the time-average of $\th$ 
during $\D \t$.  
Considering Eq.(\ref{eq:scales}), thus, it follows that 
$\mrE ^{-A(\t, \t_0)} \sim 0$, so that, 
%%%%%%%%%%%%%%%%%%%%%%%%%%%%%%%%%%%
%%%%%%%%%%%%%%%%%%%%%%%%%%%%%%%%%%%
\Beaa
&& f(\t) \sim  -\frac{1}{n-1} \int_{\t_0}^\t \bmD (\t'') \cdot \nu \ 
\mrE^{A(\t'' , \t_0)}\  d\t''  \ \mrE^{-A(\t, \t_0)} \\
&& \qquad \qquad  \sim - \D \t_{\th}  \bmD  \cdot \nu \\
&& \qquad \qquad  \sim  \D \t_{\th} \D \t  \bmD  \cdot \bmD f\ \ .
\Eeaa
%%%%%%%%%%%%%%%%%%%%%%%%%%%%%%%%%%%
%%%%%%%%%%%%%%%%%%%%%%%%%%%%%%%%%%%
Here in the second step of estimations, we have taken into account 
the time-integral is overwhelmed by the contribution in 
 the final period of order $\D \t_\th$ due to the factor $\mrE^{A(\t'' , \t_0)} $ 
along with $\D \t_\th \ll \D \t$; in the final step of estimations, 
Eq.(\ref{eq:DnuEst}) has been used.   

Since we are considering the case $f<0$, it follows 
%%%%%%%%%%%%%%%%%%%%%%%%%%%%%%%%%%%
%%%%%%%%%%%%%%%%%%%%%%%%%%%%%%%%%%%
\[
\stackrel{\circ}{\th}<0 \iff
|f(\t)| > \D \t_{\th} \D \t  |\bmD  \cdot \bmD f| \ \ .
\]
%%%%%%%%%%%%%%%%%%%%%%%%%%%%%%%%%%%
%%%%%%%%%%%%%%%%%%%%%%%%%%%%%%%%%%%
We have thus found out   the following fact: 
$f$ and $\bmD  \cdot \bmD f$ are much more important than 
$\frac{df}{d\t}$ in the later phase of gravitational contractions and 
that when $f$ is negative and  $|f(\t)|$ is  large enough to overwhelm  
$|\bmD  \cdot \bmD f|$, the contraction rate gets larger.

Now we turn to  Eq.(\ref{eq:Dth-dotC}), which  with the help of 
Eq.(\ref{eq:Rdotdot-dot}) can be recast as
%%%%%%%%%%%%%%%%%%%%%%%%%%%%%%%%%%%%
%%%%%%%%%%%%%%%%%%%%%%%%%%%%%%%%%%%%
\Bea
&& \(\frac{d \th}{d\t}\)^\circ 
= 2 \(\frac{\th^2}{n-1} + \s \cdot \s \) f - 2\th \frac{df}{d\t} 
\nonumber \\
&& \qquad  \qquad 
-\frac{2 \th}{n-1} \bmD \cdot \nu -2 \s \cdot \bmD\nu
-   \(R_{{}^\Dot {}^\Dot}\)^\circ
\ \ ,
\label{eq:Dth-dotC2} 
\Eea
%%%%%%%%%%%%%%%%%%%%%%%%%%%%%%%%%%%%
%%%%%%%%%%%%%%%%%%%%%%%%%%%%%%%%%%%%
where we have set $\om_{ab} \equiv 0$ by 
imposing the ``twist-less" condition Eq.(\ref{eq:omega-condition2}).  

Let us search for the condition for $ \(\frac{d \th}{d\t}\)^\circ <0 $. 
In later  phase of the geodesic congruence contractions, the threshold condition 
 $ \(\frac{d \th}{d\t}\)^\circ =0 $ becomes
%%%%%%%%%%%%%%%%%%%%%%%%%%%%%%%%%%%
%%%%%%%%%%%%%%%%%%%%%%%%%%%%%%%%%%%
\Bea
&&   \frac{df}{d\t}+  \(\frac{|\th|}{n-1} + \frac{\s \cdot \s}{|\th|} \) f 
\nonumber \\ 
&& \qquad 
= -\frac{ 1}{n-1} \bmD \cdot \nu +  \frac{1}{|\th|} \s \cdot \bmD\nu
+ \frac{1}{2|\th|}  \(R_{{}^\Dot {}^\Dot}\)^\circ \ .
\label{eq:border2}
\Eea  
%%%%%%%%%%%%%%%%%%%%%%%%%%%%%%%%%%%
%%%%%%%%%%%%%%%%%%%%%%%%%%%%%%%%%%%
It is quite impressive that Eq.(\ref{eq:border2}) is qualitatively similar to 
Eq.(\ref{eq:border1}) as differential equations. Looking at the first two equations 
in Eq.(\ref{eq:th-s-om-dotC}), one reasonable estimation is 
$|\th| \gg |\s_{ab}|$ as far as this inequality holds initially. 
Provided that the strong energy condition is preserved, the value of  
$|\(R_{{}^\Dot {}^\Dot}\)^\circ|$ should not be very large and it is a reasonable 
estimation that $|\th| \gg  |\(R_{{}^\Dot {}^\Dot}\)^\circ| $. 
Thus we reach almost the same conclusion that  
$f$ and $\bmD  \cdot \bmD f$ are  important 
 in the later phase of gravitational contractions and 
that, when $f$ is negative,   the large $|f(\t)|$ 
(compared to $|\bmD  \cdot \bmD f|$) gives rise to 
 the large contraction rate.

\section{\label{sec:Summary} Summary and discussions}

In this paper, we have attempted to construct a mathematical framework for  
analyzing the later stages of gravitational contractions in terms of 
the focusing properties of  timelike geodesic congruences. 

The framework has been  formed as 
a combination of a key equation,  which relates the geometry variations 
to the geodesic shifts, and 
a set of equations, which relate   
the changes in the geodesic convergence properties 
to  these geodesic shifts.
Then, what  we have constructed can be viewed as 
 a set of equations 
describing  the linear response of the 
geodesic convergence properties  to  arbitrary 
geometry variations.
Since  the geometry variations can be arbitrary  and their origins  
 need not be specified, we might be able to  
use them as probes to investigate 
the gravitational contraction processes in the present framework. 

It has then turned out that the equations get  simplified drastically in the case of 
the conformal variations.
We have then studied the latest phase of gravitational contractions 
in the case of conformal geometry variations, and have found out that 
in the final stage,  $f$ and $\bmD  \cdot \bmD f$ are much more important than 
$\frac{df}{d\t}$, and furthermore  
that the contraction rate gets larger
when $f$ is negative and  $|f|$ is  large enough to overwhelm  
$|\bmD  \cdot \bmD f|$.

It shall be fruitful to change the classes of geometry variations  and 
see the differences in the focusing properties of geodesics, as has been done 
to some degree  in 
Sec.\ref{sec:Conformal} by restricting the class to the one of the 
conformal variations and 
in Sec.\ref{sec:Focusing} by further restricting  
the class  to the one satisfying the strong energy condition.  

  The fact that the case of the conformal variations has resulted in 
 enormous simplifications in equations   might imply  that the 
non-conformal components of geometry variations contain important information on gravitational contractions which require further investigations. On the other hand, however, it means that we can make full use of the simplifications for the case of the conformal variations, which itself is important and intriguing.

One possible application related to  the conformal geometry variations might be  
the application to the black-hole spacetimes, where 
the event horizon can be characterized 
as the set of the zero-points of the conformal factor in the metric by choosing an appropriate system of coordinates.  Then the conformal geometry variations in this case 
correspond to the various deformations of the event horizon so that investigating 
their influence on the convergence properties of the geodesic congruence might be of 
significance. 

As a more  mathematically oriented  application of  the conformal variations, 
there are  issues regarding the properties of the conformal mappings. 
Studying  mathematical properties of the Penrose diagrams are a typical example
 in this category. 

It is expected that the present framework serves as a tool to understand more 
about the late stages of gravitational contractions.

\section*{Acknowledgement}
The author would like to  thank R. Miwa for valuable arguments and comments 
 helpful for completing   the present work.

\appendix

%%%%%%%%%%%%%%%%%%%%%%%%%%%%%%%%%%%%%%%%%
%%%%%%%%%%%%%%%%%%%%%%%%%%%%%%%%%%%%%%%%%
%APPENDIX A
%%%%%%%%%%%%%%%%%%%%%%%%%%%%%%%%%%%%%%%%%
%%%%%%%%%%%%%%%%%%%%%%%%%%%%%%%%%%%%%%%%%
\section{\label{app:formulas}Useful formulas}

Below we enumerate  useful formulas frequently used in the body of the paper.

Some among them are formulas regarding the $((n-1)+1)$-decomposition of 
several quantities.  For their derivations, 
it is helpful to use the relation Eq.(\ref{eq:hab}) and its 
variations, such as 
\[
{\d_a}^b= {h_a}^b - \xi_a \xi^b\ \ .
\]

With the help of  these relations,  any vector $u_a$ can be decomposed as 
%%%%%%%%%%%%%%%%%%%%%%%%%%%%%%%%%%%%%
%%%%%%%%%%%%%%%%%%%%%%%%%%%%%%%%%%%%%
\Beq
u_a  = \ul{u}_a  - u^\Dot \xi_a \ \ , 
\label{eq:decovec}
\Eeq
%%%%%%%%%%%%%%%%%%%%%%%%%%%%%%%%%%%%
%%%%%%%%%%%%%%%%%%%%%%%%%%%%%%%%%%%%
where $u^\Dot = u_b \xi^b$ (see Sec.{\ref{subsec:notations}} for notations employed in this paper).

For a scalar function $f$, then, the 
application of  Eq.(\ref{eq:decovec})  to a vector $\nb_a f$  yields  
%%%%%%%%%%%%%%%%%%%%%%%%%%%%%%%%%%%%%
%%%%%%%%%%%%%%%%%%%%%%%%%%%%%%%%%%%%%
\Beq
\nb_a f = \bmD_a f -  \frac{df}{d\t} \xi_a\ \ , 
\label{eq:nablaF}
\Eeq
%%%%%%%%%%%%%%%%%%%%%%%%%%%%%%%%%%%%
%%%%%%%%%%%%%%%%%%%%%%%%%%%%%%%%%%%%
where $\bmD_a$ is  the spatial  
derivative operator induced from $\nb_a$ (Eq.(\ref{eq:bmD})). 

For notational brevity, it is convenient to introduce
%%%%%%%%%%%%%%%%%%%%%%%%%%%%%%%%%%
\[
\nbbr_a \equiv \bmD_a -   \xi_a \ddt \ \ .
\]
%%%%%%%%%%%%%%%%%%%%%%%%%%%%%%%%%%
It is important to note that $\nbbr_a$ coincides with 
the standard $\nb_a$ only  when the former  is applied to functions 
(Eq.(\ref{eq:nablaF})).

For a general spacetime vector $u^a$, it follows
%%%%%%%%%%%%%%%%%%%%%%%%%%%%%%%%%%%%%
%%%%%%%%%%%%%%%%%%%%%%%%%%%%%%%%%%%%%
\Beq
\nabla_b u^a = \nbbr_b \ul{u}^a   
+ \xi^a (\ul{u}B)_{b}   - \xi^a \nbbr_b u^\Dot - u^\Dot {B^a}_b \ \ . 
\label{eq:nablaU}
\Eeq
%%%%%%%%%%%%%%%%%%%%%%%%%%%%%%%%%%%%
%%%%%%%%%%%%%%%%%%%%%%%%%%%%%%%%%%%%
For a spatial vector $\ul{u}^a$, then, Eq.(\ref{eq:nablaU}) reduces to 
%%%%%%%%%%%%%%%%%%%%%%%%%%%%%%%%%%%%%
%%%%%%%%%%%%%%%%%%%%%%%%%%%%%%%%%%%%%
\Beq
\nabla_b \ul{u}^a = \nbbr_b \ul{u}^a + \xi^a (\ul{u}B)_{b} \ \ .   
\label{eq:nablaUbar}
\Eeq
%%%%%%%%%%%%%%%%%%%%%%%%%%%%%%%%%%%%
%%%%%%%%%%%%%%%%%%%%%%%%%%%%%%%%%%%%
It is obvious that the contravariant index $a$ in Eq.(\ref{eq:nablaU}) and 
Eq.(\ref{eq:nablaUbar}) can freely be lowered and the corresponding formulas with  
the covariant index $a$ also follow.

Let  $\g_{ab}$ be  a symmetric covariant tensor.   Then it follows 
%%%%%%%%%%%%%%%%%%%%%%%%%%%%%%%%%%%%%
%%%%%%%%%%%%%%%%%%%%%%%%%%%%%%%%%%%%%
\Bea
&& \g_{ab}  
= {\ul{\g}\ }_{ab}  -  2 \xi_{\ol{a}} \ {\ul{\g}\ }_{\ol{b}} 
+ \gddot \xi_a \xi_b  \ \ ,  \nonumber \\
&& \qquad   
= {\ul{\g}\ }_{ab}  -  2 \xi_{\ol{a}} \ {\g}_{\ol{b}} 
- \gddot \xi_a \xi_b \ \ , 
\label{eq:decogamma}
\Eea
%%%%%%%%%%%%%%%%%%%%%%%%%%%%%%%%%%%%
%%%%%%%%%%%%%%%%%%%%%%%%%%%%%%%%%%%%
where ${\ul{\g}\,}{}_b$ is the shorthand notation for 
the spatial part of 
$\g_{b\, {}^\Dot}=\g_{b c} \xi^c $.
We also note the difference between     ${\ul{\g}\,}{}_b$  and 
$\g_b$  and the sign difference in the first and the second lines.

Taking the derivative $\nb_a$ on both sides of Eq.(\ref{eq:decogamma}) with 
some straightforward computations,   
one can obtain  another formula for  $\g_{ab}$, 
%%%%%%%%%%%%%%%%%%%%%%%%%%%%%%%%%%%%%
%%%%%%%%%%%%%%%%%%%%%%%%%%%%%%%%%%%%%
\Bea
&& \nb_b \g_{cd}  = \nbbr_b {\ul{\g}\ }_{cd}  
+  2 \xi_{\ol{c}} \ ({\ul{\g}\ }  B )_{\ol{d}\ b}  
  -2 \ul{\g}_{\ol{c}} \  B_{\ol{d}\ b} 
-2 \xi_{\ol{c}} \  \nbbr_b {\ul{\g}}_{\ol{d}}  \nonumber \\
&& \qquad -2 (\ul{\g \ } B )_b \xi _c \xi_d
+ \nbbr_b \gddot \  \xi_c \xi_d 
+ 2  \gddot \xi_{\ol{c}} B_{\ol{d} \ b}  \ \ ,    
\label{eq:nbgamma}
\Eea
%%%%%%%%%%%%%%%%%%%%%%%%%%%%%%%%%%%%
%%%%%%%%%%%%%%%%%%%%%%%%%%%%%%%%%%%%
which can most easily be derived by applying Eq.(\ref{eq:nablaU}) to $u_a u_b$.

In particular, for a symmetric spatial tensor $ {\ul{\g}\ }_{ab}$, 
Eq.(\ref{eq:nbgamma}) simplifies to 
%%%%%%%%%%%%%%%%%%%%%%%%%%%%%%%%%%%%%
%%%%%%%%%%%%%%%%%%%%%%%%%%%%%%%%%%%%%
\Bea
&& \nb_b {\ul{\g}\  }_{cd}  = \nbbr_b {\ul{\g}\ }_{cd}  
+  2 \xi_{\ol{c}} \ ({\ul{\g}\ } B )_{\ol{d}\ b}  \ \ .    
\label{eq:nbgamma2}
\Eea
%%%%%%%%%%%%%%%%%%%%%%%%%%%%%%%%%%%%
%%%%%%%%%%%%%%%%%%%%%%%%%%%%%%%%%%%%

There are some useful formulas for the second derivatives of a  function $f$;  
%%%%%%%%%%%%%%%%%%%%%%%%%%%%%%%%%%%
%%%%%%%%%%%%%%%%%%%%%%%%%%%%%%%%%%%
\Beq
\ddt \bmD_a f = \bmD_a \frac{df}{d\t} - {B^c}_a \bmD_c f \ \ ,
\label{eq:dD-Dd}
\Eeq
%%%%%%%%%%%%%%%%%%%%%%%%%%%%%%%%%%%
%%%%%%%%%%%%%%%%%%%%%%%%%%%%%%%%%%%
and 
%%%%%%%%%%%%%%%%%%%%%%%%%%%%%%%%%%%
%%%%%%%%%%%%%%%%%%%%%%%%%%%%%%%%%%%
\Beq
\nb_a \nb_b f = \bmD_a \bmD_b f   - B_{ba} \frac{df}{d\t}
-2\xi_{\ol{a}} \ddt \bmD_{\ol{b}} f + \xi_a \xi_b \frac{d^2 f}{d\t^2}
   \ \ , 
\label{eq:nbnb-DD}
\Eeq
%%%%%%%%%%%%%%%%%%%%%%%%%%%%%%%%%%%
and also
%%%%%%%%%%%%%%%%%%%%%%%%%%%%%%%%%%%
\Beq
\bmD_a \bmD_b f = {h^c}_a {h^d}_b \nb_c \nb_d f 
+ B_{ba} \frac{df}{d\t}\ \ .
\label{eq:bmDbmD}
\Eeq
%%%%%%%%%%%%%%%%%%%%%%%%%%%%%%%%%%%%%%
%%%%%%%%%%%%%%%%%%%%%%%%%%%%%%%%%%%%%

By the contraction of Eq.(\ref{eq:nbnb-DD}), we get
%%%%%%%%%%%%%%%%%%%%%%%%%%%%%%%%%%%
%%%%%%%%%%%%%%%%%%%%%%%%%%%%%%%%%%%
\Beq
\D f = \bmD \cdot \bmD  f - \th \frac{df}{d\t} - \frac{d^2 f}{d\t^2}\ \ .
\label{eq:Delta-Delta}
\Eeq
%%%%%%%%%%%%%%%%%%%%%%%%%%%%%%%%%%%
%%%%%%%%%%%%%%%%%%%%%%%%%%%%%%%%%%%
On the other hand, by the anti-symmetrization of Eq.(\ref{eq:nbnb-DD}), 
we get
%%%%%%%%%%%%%%%%%%%%%%%%%%%%%%%%%%%
%%%%%%%%%%%%%%%%%%%%%%%%%%%%%%%%%%%
\Beq
(\bmD_a  \bmD_b - \bmD_b  \bmD_a)  f 
=  -2 \om_{ab} \frac{df}{d\t} \ \ ,
\label{eq:bmD-Torsion}
\Eeq
%%%%%%%%%%%%%%%%%%%%%%%%%%%%%%%%%%%
%%%%%%%%%%%%%%%%%%%%%%%%%%%%%%%%%%%
implying that the  induced covariant derivative $\bmD_a$ is torsion-free {\sl iff}  
$\om_{ab}=0$.  
Thus the following statements are mutually equivalent:
%%%%%%%%%%%%%%%%%%%%%%%%%%%%%%
\BD
\item{(1)}  $\om_{ab}=0$ $\iff$ $\nb_a \xi_b = \nb_b \xi_a$ 
                    $\iff$  $\bmD_a  \bmD_b f = \bmD_b  \bmD_a  f$.  
\item{(2)}  The timelike geodesic congruence $\C$ in question 
                   is hypersurface orthogonal, i.e. $\C$ can be foliated by 
                    smooth $(n-1)$-dimensional orthogonal sections.
\item{(3)}  $B_{ab}$  coincides with  the extrinsic curvature $K_{ab}$ 
       for  the smooth orthogonal sections of  $\C$.
\ED
%%%%%%%%%%%%%%%%%%%%%%%%%%%%%%%

We here pay attention to a useful formula including the $\circ$-derivative.
By taking the $\circ$-derivative on both sides of 
$h_{ab}= g_{ab} + \xi_a \xi_b$ or 
$h^{ab}= g^{ab} + \xi^a \xi^b$
and using Eq.(\ref{eq:g-dot}),  
Eq.(\ref{eq:g-dot2}),  Eq.(\ref{eq:nu2}), Eq.(\ref{eq:decovec}) and
Eq.(\ref{eq:decogamma}), we get 
%%%%%%%%%%%%%%%%%%%%%%%%%%%%%%%%
%%%%%%%%%%%%%%%%%%%%%%%%%%%%%%%%
\Bea
&& \stackrel{\circ}{h}{}_{ab}= \ul{\g}{}_{ab} + 2 \xi_{\ol{a}}\, \nu_{\ol{b}}
\ \ ,  \nonumber \\
&&
\stackrel{\circ}{h}{}^{ab}= - \ul{\g}{}_{ab} 
+ 2 \xi_{\ol{a}}\,  \( \nu_{\ol{b}} + {\ul{\g}\, }_{\ol{b}} \)
\ \ .
\label{eq:h-dot}
\Eea
%%%%%%%%%%%%%%%%%%%%%%%%%%%%%%%%
%%%%%%%%%%%%%%%%%%%%%%%%%%%%%%%%
Taking the $\circ$-derivative on both sides of 
${h_a}^b ={h_a}^c {h_c}^b $ and using Eq.(\ref{eq:h-dot}), 
it is straightforward
 to show that
%%%%%%%%%%%%%%%%%%%%%%%%%%%%%%%%
%%%%%%%%%%%%%%%%%%%%%%%%%%%%%%%%
\Beq
 \stackrel{\circ}{h}{{}_a}^b= 0\ \ .
\label{eq:h-dot-null}
\Eeq
%%%%%%%%%%%%%%%%%%%%%%%%%%%%%%%%
%%%%%%%%%%%%%%%%%%%%%%%%%%%%%%%%

\section{\label{app:FRWapp}Basic facts on the Friedmann-Robertson-Walker model}

We here summarize the basic points on the Friedmann-Robertson-Walker (FRW) 
model discussed in Sec.\ref{subsec:FRW}.

The FRW spacetime is a spacetime  foliated by   
maximally symmetric (i.e. homogeneous and isotropic) spatial hypersurfaces.  
The metric for the FRW spacetime is then given  as
%%%%%%%%%%%%%%%%%%%%%%%%%%%%%%%%%%%
%%%%%%%%%%%%%%%%%%%%%%%%%%%%%%%%%%%
\Bea
&& ds^2
    = -dt^2 +a^2(t) \left( F^2(r) dr^2 + r^2 d\vartheta^2+ r^2\sin^2\vartheta 
d\varphi^2 \right)
\ \ , 
\nonumber \\
&& F(r) := \frac{1}{\sqrt{1-kr^2}} \qquad (k= -1,0,1)\ \ .
\label{eq:FRWmetric}
\Eea
%%%%%%%%%%%%%%%%%%%%%%%%%%%%%%%%%%%
%%%%%%%%%%%%%%%%%%%%%%%%%%%%%%%%%%%
The world-lines of  observers ``standing still" relative 
to the  system of coordinates, Eq.(\ref{eq:FRWmetric}),  
is described by  
$x^a(t) = {}^T (t\   r_0\  \vartheta_0\  \varphi_0 )$ with 
$r_0$, $\vartheta_0$ and $\varphi_0$ 
being  some constants.  They are easily shown to be timelike geodesics. 
Let us call these geodesics  {\it standard geodesics} compatible with 
 the system of   coordinates, Eq.(\ref{eq:FRWmetric}). 

 The spatial part of the coordinate system ($\{ r, \th, \phi \}$), thus,
 is interpreted  as the collection of the ``names" of  the  observers  free-falling  
along  each standard geodesic. 
We also note that the time-coordinate $t$ coincides with the proper-time $\t$ for the 
free-falling observer  along a standard geodesic. 
Hereafter each standard geodesic is assumed to be parametrized by 
the proper-time $\t$, coinciding with $t$. 
Based on these facts, we see that   
the coordinate system $\{t, r, \vartheta, \varphi \}$ 
forms a {\it co-moving} coordinate system.

It is sometimes convenient to shift from the coordinates 
$\{t , r, \vartheta, \varphi \}$ to the local  
pseudo-ortho-normal frame of coordinates $\{ \Th^0, \Th^1, \Th^2, \Th^3 \}$ 
defined as  
%%%%%%%%%%%%%%%%%%%%%%%%%%%%%%%%%%%
%%%%%%%%%%%%%%%%%%%%%%%%%%%%%%%%%%%
\Bea
&& \Th^0 = dt\ \ , \ \ \Th^1= a(t) F(r) dr\ \ , 
\ \ \Th^2 =a(t) r d \vartheta \ \ ,  
\nonumber \\
&&  \Th^3 = a(t) r \sin \vartheta d\varphi\ \ ,
\label{eq:ONframe}
\Eea
%%%%%%%%%%%%%%%%%%%%%%%%%%%%%%%%%%%
%%%%%%%%%%%%%%%%%%%%%%%%%%%%%%%%%%%
which makes the expression for $ds^2$ simply 
%%%%%%%%%%%%%%%%%%%%%%%%%%%%%%%%%%%
%%%%%%%%%%%%%%%%%%%%%%%%%%%%%%%%%%%
\Beq
ds^2 = - (\Th^0)^2 + (\Th^1)^2 +(\Th^2)^2 +(\Th^3)^2\ \ .
\label{eq:metricCartan}
\Eeq
%%%%%%%%%%%%%%%%%%%%%%%%%%%%%%%%%%%
%%%%%%%%%%%%%%%%%%%%%%%%%%%%%%%%%%%

Let $\xi^a$ be the tangent vector of a standard geodesic and let 
$h_{ab}$ be given by Eq.(\ref{eq:hab}).     
We then see that  
$\xi^a={}^T (1\ 0\ 0\ 0)$,  $\xi_a={}^T (-1\ 0\ 0\ 0)$ and  
 $h_{ab}={\rm diag} (0\ 1\ 1\ 1)$  
in the $\{ \Th^a \}$ frame.

Following the  standard procedures for  
differential forms~\cite{Franders, Nakahara}, then, 
we first get the expression for the connection 1-forms ${\Lambda^a}_b$ as
%%%%%%%%%%%%%%%%%%%%%%%%%%%%%%%%%%%
%%%%%%%%%%%%%%%%%%%%%%%%%%%%%%%%%%%
\Bea
&& {\Lambda^0}_k = {\Lambda^k}_0 
= \frac{\dot{a}}{a}\,  \Th^k\ \ (k=1,2,3)\ \ , 
\nonumber \\
&& {\Lambda^1}_m=-{\Lambda^m}_1 
= - \frac{1}{arF}\, \Th^m \ \ (m=1,2) \ \ , 
\nonumber \\
&& {\Lambda^2}_3=-{\Lambda^3}_2 = - \frac{1}{ar}\, \cot \vartheta \, \Th^3 \ \ .
\label{eq:connectionFRW}
\Eea
%%%%%%%%%%%%%%%%%%%%%%%%%%%%%%%%%%%
%%%%%%%%%%%%%%%%%%%%%%%%%%%%%%%%%%%
From Eq.(\ref{eq:connectionFRW}), 
 we can derive the expression for   the Riemann curvature as  
%%%%%%%%%%%%%%%%%%%%%%%%%%%%%%%%%%%
%%%%%%%%%%%%%%%%%%%%%%%%%%%%%%%%%%%
\Bea
&& {R^0}_{k0k} =\frac{\ddot{a}}{a}\ \  (k=1,2,3) \ \ , 
\nonumber \\
&&  {R^1}_{m1m} = \(\frac{\dot{a}}{a}\)^2 + \frac{F'}{a^2rF^3}\ \ 
(m=1,2) \ \ , 
\nonumber \\
&& {R^2}_{323}= \(\frac{\dot{a}}{a}\)^2 - \frac{1-F^2}{a^2r^2F^2}\ \ , 
\label{eq:Riemann-Cartan}
\Eea
%%%%%%%%%%%%%%%%%%%%%%%%%%%%%%%%%%%
%%%%%%%%%%%%%%%%%%%%%%%%%%%%%%%%%%%
with the other independent components being zero.
From Eq.(\ref{eq:Riemann-Cartan}), we get useful coordinate-independent  relations 
%%%%%%%%%%%%%%%%%%%%%%%%%%%%%%%%%%
%%%%%%%%%%%%%%%%%%%%%%%%%%%%%%%%%%
\Beq
R_{abc^\Dot}= 2 \frac{\ddot{a}}{a}h_{\ul{a}c} \xi_{\ul{b}}\ \ , \ \ 
R_{a^\Dot  b^\Dot}   =-\frac{\ddot{a}}{a} h_{ab}\ \ .
\label{eq:Rabcdot}
\Eeq
%%%%%%%%%%%%%%%%%%%%%%%%%%%%%%%%%%
%%%%%%%%%%%%%%%%%%%%%%%%%%%%%%%%%%

Then the Ricci curvature becomes
%%%%%%%%%%%%%%%%%%%%%%%%%%%%%%%%%%%
%%%%%%%%%%%%%%%%%%%%%%%%%%%%%%%%%%%
\Bea
&& R_{00} =-3 \frac{\ddot{a}}{a}\ \ , 
\nonumber \\
&&  R_{11} = 2 \(\frac{\dot{a}}{a}\)^2 
                         + \frac{\ddot{a}}{a} +  \frac{2F'}{a^2rF^3}\ \ , 
\nonumber \\
&& R_{22} = R_{33} =  2 \(\frac{\dot{a}}{a}\)^2 
                         + \frac{\ddot{a}}{a} +  \frac{rF' + F^3 -F}{a^2r^2F^3}\ \ ,
\label{eq:Ricci-Cartan}
\Eea
%%%%%%%%%%%%%%%%%%%%%%%%%%%%%%%%%%%
%%%%%%%%%%%%%%%%%%%%%%%%%%%%%%%%%%%
with the other  components being zero. From Eq.(\ref{eq:Ricci-Cartan}),  we 
get other useful coordinate-independent relations 
%%%%%%%%%%%%%%%%%%%%%%%%%%%%%%%%%%%
%%%%%%%%%%%%%%%%%%%%%%%%%%%%%%%%%%%
\Beq
R_{a^\Dot}=3\frac{\ddot{a}}{a} \xi_a\ \ , \ \ 
R_{{}^\Dot  {}^\Dot} = -3\frac{\ddot{a}}{a}\ \ .
\label{eq:Rdotdot}
\Eeq
%%%%%%%%%%%%%%%%%%%%%%%%%%%%%%%%%%%
%%%%%%%%%%%%%%%%%%%%%%%%%%%%%%%%%%%
Finally the scalar curvature becomes
%%%%%%%%%%%%%%%%%%%%%%%%%%%%%%%%%%%
%%%%%%%%%%%%%%%%%%%%%%%%%%%%%%%%%%%
\Beq
 R =6  \frac{\ddot{a}}{a} + 6 \(\frac{\dot{a}}{a}\)^2 
                         + 2 \frac{2rF' + F^3 -F}{a^2r^2F^3}\ \ .
\label{eq:scalar-Cartan}
\Eeq
%%%%%%%%%%%%%%%%%%%%%%%%%%%%%%%%%%%
%%%%%%%%%%%%%%%%%%%%%%%%%%%%%%%%%%%
For reference, let us write down the expression for the Einstein tensor also;
%%%%%%%%%%%%%%%%%%%%%%%%%%%%%%%%%%%
%%%%%%%%%%%%%%%%%%%%%%%%%%%%%%%%%%%
\Bea
&& G_{00} =3 \(\frac{\dot{a}}{a}\)^2 + \frac{2rF' + F^3 -F}{a^2r^2F^3} \ \ , 
\nonumber \\
&&  G_{11} =  - \(\frac{\dot{a}}{a}\)^2 -2 \frac{\ddot{a}}{a} 
                                 +  \frac{1-F^2}{a^2r^2F^2}\ \ , 
\nonumber \\
&& G_{22} = G_{33} =  - \(\frac{\dot{a}}{a}\)^2 -2 \frac{\ddot{a}}{a}\ \ ,
\label{eq:Einstein-Cartan}
\Eea
%%%%%%%%%%%%%%%%%%%%%%%%%%%%%%%%%%%
%%%%%%%%%%%%%%%%%%%%%%%%%%%%%%%%%%%
with the other  components being zero.

%%%%%%%%%%%%%%%%%%%%%%%%%%%%%%%%%%%%%
%REFERENCES
%%%%%%%%%%%%%%%%%%%%%%%%%%%%%%%%%%%%%%%%%%%%%%%%%%%%%%%%%%%%%%%%%


\begin{thebibliography}{99}
\bibitem{footnote1}
Einstein's summation convention is adopted throughout this paper. We shall also 
adopt a suitable system of units in which the light velocity $c$ becomes  $c=1$.
\bibitem{footnote2}
For notational convenience, we shall assume 
below that  the intervals such as  $(s_1, s_2)$ here and  $\Lambda$ in 
Sec.{\ref{subsec:1-parameter}} contain $0$ whenever necessary.
\bibitem{footnote3} 
Here the concepts of ``orthogonal" and 
``parallel" are naturally judged in the original geometry $(\M , g)$.
\end{thebibliography}
\end{document}